\documentclass[12pt]{article}
\pdfoutput=1
\usepackage[totalwidth=6.5in,totalheight=9in]{geometry}
\usepackage[centertags]{amsmath}
\usepackage[square,comma,sort&compress,numbers]{natbib}
\usepackage{epsfig}
\usepackage{amsfonts}
\usepackage{amssymb}
\usepackage{slashed}
\usepackage{mathrsfs}
\usepackage{bbm}
\usepackage{array,multirow}
\usepackage{graphicx}
\usepackage{caption}
\usepackage{subcaption}
\usepackage{xcolor}
\usepackage{setspace}
\usepackage{comment}
\usepackage{enumitem}
\usepackage{hyperref}

\newcommand{\ra}{\rightarrow}

\newcommand{\lra}{\leftrightarrow}

\newcommand{\<}{\langle}
\renewcommand{\>}{\rangle}
\newcommand{\be}{\begin{equation}}
\newcommand{\ee}{\end{equation}}
\newcommand{\ba}{\begin{aligned}}
\newcommand{\ea}{\end{aligned}}
\newcommand{\benn}{\begin{equation*}}
\newcommand{\eenn}{\end{equation*}}
\newcommand{\bi}{\begin{itemize}}  
\newcommand{\ei}{\end{itemize}}
\newcommand{\bpm}{\begin{pmatrix}}
\newcommand{\epm}{\end{pmatrix}}

\newcommand{\p}{\partial}

\newcommand{\De}{\Delta}

\newcommand{\vep}{\varepsilon}
\newcommand{\G}{\Gamma}

\newcommand{\Ccal}{{\mathcal C}}

\newcommand{\Ecal}{{\mathcal E}}

\newcommand{\Ncal}{{\mathcal N}}
\newcommand{\Ocal}{{\mathcal O}}

\newcommand{\gap}{\textrm{gap} }

\newcommand{\fr}{\frac}
\newcommand{\tfr}{\tfrac}
\newcommand{\half}{\frac{1}{2}}

\newcommand{\comm}[2]{[#1,#2]}

\newcommand{\Chat}{\widehat{C}}
\newcommand{\nmax}{{n_{\max}}}
\newcommand{\nmin}{{n_{\min}}}

\newcommand{\Lmax}{{\ell_{\max}}}

\renewcommand{\;}{\mspace{1mu}}

\numberwithin{equation}{section}

\begin{document}

\begin{titlepage}

\noindent\makebox[\dimexpr\linewidth-1cm\relax][r]{LA-UR-26-25971}

\setcounter{page}{0}

\begin{center}

\vspace*{2cm}

{\bf \Large The EEC-Hedron:} \\
\vspace{0.2cm}
{\bf \Large Positivity Bounds on Energy Correlators}

\vspace{0.5cm}

{\bf Bianka Me\c{c}aj$^1$, Ian Moult$^2$, Hidde Stoffels$^3$, Matthew T.\ Walters$^3$, Yuan Xin$^{4,5,6}$}

\vspace{0.5cm}

{\it $^1$ Theoretical Division, Los Alamos National Laboratory, Los Alamos, NM 87545, USA} \\
{\it $^2$ Department of Physics, Yale University, New Haven, CT 06511, USA} \\
{\it $^3$ Maxwell Institute for Mathematical Sciences, Department of Mathematics, \\
Heriot-Watt University, Edinburgh EH14, UK} \\
{\it $^4$ Center for Mathematics and Interdisciplinary Sciences, Fudan University, \\
Shanghai, 200433, China}\\
{\it $^5$ Shanghai Institute for Mathematics and Interdisciplinary Sciences (SIMIS), \\
Shanghai, 200433, China}\\
{\it $^6$ Department of Physics, Carnegie Mellon University, Pittsburgh, PA 15213, USA}\\

\end{center}

\vspace{1cm}

\begin{abstract}

We study the constraints placed by energy positivity on general unitary quantum field theories. We show that the coefficients in the partial wave expansion of energy-energy correlators all must lie within a bounded region of parameter space, which we call the ``EEC-hedron''. For the particular case of conformal field theories, these positivity constraints on energy correlators lead to bounds on the allowed values of coefficients in the operator product expansion. We demonstrate this approach in the 3d Ising and $O(2)$ CFTs, deriving new bounds on OPE coefficients in both theories.

\end{abstract}

\end{titlepage}

\tableofcontents


\section{Introduction}
\label{sec:Intro}

The basic principles of unitarity and locality impose significant constraints on the space of possible quantum field theories. While ultimately encoded in underlying axioms, it has proven extremely powerful to phrase these as statements about the behavior of physical observables, the specific choice of which depends on the properties of the QFT of interest. 

In a theory with particle excitations, the S-matrix bootstrap encodes these fundamental axioms as analyticity and crossing symmetry of scattering amplitudes~\cite{Kruczenski:2022lot}. In conformal field theories, the conformal bootstrap encodes them in unitarity and crossing for correlation functions of local operators~\cite{Poland:2018epd,Hartman:2022zik,Rychkov:2023wsd}. In a similar spirit, approaches based on operator algebras use basic inequalities of entanglement~\cite{Casini:2022rlv,Faulkner:2022mlp}. In the case that one has access to all observables of a given type, it is believed that in principle all such methods are equivalent. However, in practice these different types of constraints provide complementary windows into different aspects of the underlying physics.

One powerful set of observables are correlation functions of the energy flow operator (averaged null energy operator)~\cite{Sterman:1975xv,Basham:1977iq,Basham:1978bw,Basham:1978zq,Basham:1979gh,Sveshnikov:1995vi,Tkachov:1995kk,Korchemsky:1999kt}
\be
\Ecal(\hat{n}) \equiv \lim_{r\ra\infty} r^{d-2} \int_0^\infty dt \, \hat{n}^i \; T_{0i}(t,r\hat{n}),
\label{eq:EdefNaive}
\ee
which measures the total energy flux in the direction specified by the unit vector $\hat{n}$. One of the useful features of energy correlator observables is that they exist in both gapped theories with an S-matrix and CFTs. More generally, they are expected to exist even in the case of gravity coupled to a CFT, where one has neither an S-matrix nor local correlators~\cite{Herrmann:2024yai,Chicherin:2025keq}. However, as compared to the case of the S-matrix or correlation functions of local operators, they are much less explored (see~\cite{Moult:2025nhu} for a recent review).

Perhaps the most important constraint on energy correlators is the Averaged Null Energy Condition (ANEC),
\be
\<\psi|\Ecal(\hat{n})|\psi\> \geq 0 \textrm{ for all } |\psi\>,
\label{eq:ANEC}
\ee
which has been rigorously proven in unitary relativistic QFTs, both from causality~\cite{Hartman:2016lgu} and monotonicity of entanglement entropy~\cite{Faulkner:2016mzt}. The ANEC has been used to derive a number of powerful results, including the ``conformal collider'' bounds on the anomaly coefficients $a$ and $c$ in conformal field theories~\cite{Hofman:2008ar}, constraints on scaling dimensions~\cite{Cordova:2017dhq} and OPE coefficients~\cite{Hofman:2009ug,Chowdhury:2017vel,Cordova:2017zej}, bounds on transport coefficients~\cite{Delacretaz:2018cfk}, and the $a$- and $c$-theorems for renormalization group flows~\cite{Hartman:2023qdn,Hartman:2023ccw,Hartman:2024xkw}.

For the case of conformal field theories, there has been significant recent progress in determining the constraints placed by locality and unitarity on energy correlators~\cite{Belitsky:2013ofa,Belitsky:2013bja,Belitsky:2013xxa,Kravchuk:2018htv,Cordova:2018ygx,Belin:2019mnx,Kologlu:2019bco,Kologlu:2019mfz,Chang:2020qpj,Belin:2020lsr,Korchemsky:2021okt,Korchemsky:2021htm,Caron-Huot:2022eqs,Chicherin:2023gxt,Firat:2023lbp,Chen:2024iuv,Cuomo:2025pjp,Mecaj:2025ecl,Dempsey:2025yiv}. In particular, it has been proven that energy flow operators commute~\cite{Kologlu:2019bco} (see also~\cite{Casini:2017roe,Cordova:2018ygx,Belin:2019mnx})
\be
\comm{\Ecal(\hat{n}_1)}{\Ecal(\hat{n}_2)} = 0,
\label{eq:Commute}
\ee
which suggests a much stronger version of averaged null energy positivity
\be
\boxed{\<\psi|\Ecal(\hat{n}_1) \cdots \Ecal(\hat{n}_k)|\psi\> \geq 0 \textrm{ for all } |\psi\>.}
\label{eq:MultiPoint}
\ee
This multi-point positivity also automatically follows from the definition of energy correlators in terms of S-matrix elements in theories with particle excitations~\cite{Fox:1978vw}. 
It would be very useful to rigorously prove that~\eqref{eq:MultiPoint} holds in all unitary quantum field theories.

In this paper, we take~\eqref{eq:MultiPoint} to be one of the axioms of energy correlators, in order to determine the resulting constraints on the space of consistent QFTs.\footnote{In~\cite{Belin:2026wkc} it was argued that these constraints are part of an infinite hierarchy of bounds obtained by imposing consistency between all higher-point energy correlators in the background of the same external state.} As compared to the standard ANEC~\eqref{eq:ANEC}, multi-point positivity constraints have been much less explored, with the notable exception of~\cite{Zhiboedov:2013opa}, which used~\eqref{eq:MultiPoint} to show that CFTs which saturate the conformal collider bounds must be non-interacting. The effects of multi-point positivity were recently studied in \cite{Mecaj:2025ecl,Dempsey:2025yiv} for the specific case where the external state $|\psi\>$ is created by a scalar operator. It was shown that the energy-energy correlator (EEC) can be decomposed into partial waves, which in $d=4$ takes the form
\be
\<\psi|\Ecal(\hat{n}_1) \; \Ecal(\hat{n}_2)|\psi\> = \fr{\<\psi|\Ecal(\hat{n}_1)|\psi\> \; \<\psi|\Ecal(\hat{n}_2)|\psi\>}{\<\psi|\psi\>} \sum_{\ell=0}^\infty F_\ell \; (2\ell+1) \; P_\ell(\cos\theta)\,,
\label{eq:PartialWave4d}
\ee
with coefficients $F_\ell$ which are positive and bounded
\be
\boxed{0 \leq F_\ell \leq 1.}
\label{eq:SimpleBound}
\ee

In~\cite{Mecaj:2025ecl}, the bounds~\eqref{eq:SimpleBound} on EEC partial wave coefficients were used to derive constraints on OPE coefficients in the 3d Ising CFT. However, this is an extremely weak use of positivity of the two-point correlator. The partial wave coefficients are highly correlated (in fact they are analytic in $\ell$~\cite{Dempsey:2025yiv}), such that generic combinations of $F_\ell$ satisfying~\eqref{eq:SimpleBound} do not correspond to a consistent QFT. Determining which combinations of coefficients are consistent with positivity of the full energy correlator is a standard moment problem, analogous to the positivity constraints explored extensively in the context of the analytic S-matrix bootstrap~\cite{Bellazzini:2020cot,Tolley:2020gtv,Caron-Huot:2020cmc,Arkani-Hamed:2020blm}, and was used in~\cite{Dempsey:2025yiv} to significantly constrain the structure of the EEC in $\Ncal=4$ super-Yang-Mills.

In this paper, we formulate the study of positivity of energy correlators as a moment problem. This carves out a space of allowed values for the coefficients $F_\ell$, in which the energy correlators of consistent unitary relativistic QFTs must live. In analogy with the EFT-hedron~\cite{Arkani-Hamed:2020blm} and the Hydro-hedron~\cite{Heller:2023jtd}, we term this space of physical EEC partial wave coefficients the ``EEC-hedron".

We explore the EEC-hedron for QFTs in general dimensions $d\geq 3$. For the specific case of CFTs, we combine the EEC-hedron with the conformal blocks computed in~\cite{Mecaj:2025ecl}, as well as data from the numerical conformal bootstrap~\cite{Simmons-Duffin:2016wlq,Chester:2019ifh,Liu:2020tpf,Chang:2024whx,Poland:2025ide} and fuzzy sphere regularization~\cite{Fardelli:2026zas,Dey:2026cso}, to derive novel bounds on OPE coefficients in the 3d Ising and $O(2)$ CFTs. For simplicity, we specifically consider energy correlators in the background created by a scalar source, leaving the generalization to spinning sources for future work.\footnote{The spinning case was recently studied for the one-point energy correlator $\<\psi|\Ecal(\hat{n})|\psi\>$ in~\cite{Riembau:2024tom,Riembau:2025wjc}, resulting in analogous positivity constraints on its allowed structure in consistent QFTs.}

An outline of this paper is as follows. In section~\ref{sec:EECBounds}, we review the derivation of moment positivity bounds from the study of the analytic S-matrix bootstrap and use these to derive positivity constraints on energy correlators, introducing the EEC-hedron. We consider both the case of QFTs, as relevant for real world QCD, as well as CFTs. In section~\ref{sec:IsingBounds}, we apply this approach to the case of the 3d Ising CFT, deriving both upper and lower bounds on OPE coefficients. In section~\ref{sec:O2Bounds}, we perform a similar analysis for the case of the 3d $O(2)$ CFT. In section~\ref{sec:Bootstrap}, we demonstrate how these positivity constraints can be rephrased into a form suitable for standard numerical bootstrap methods. We conclude in section~\ref{sec:Discussion} and discuss a number of future directions.


\section{Universal Bounds on Energy Correlators}
\label{sec:EECBounds}

In this section we explore constraints from multi-point ANEC positivity on generic unitary quantum field theories, extending the initial results in~\cite{Mecaj:2025ecl,Dempsey:2025yiv}. Here, and throughout the paper, we explore positivity in the simplest case of a state produced by a local scalar operator
\be
|\psi\> = |\phi(p)\> \equiv \int d^dx \; e^{-ip\cdot x} \phi(x)|0\>\;,
\label{eq:PhiDef}
\ee
where $p^\mu$ denotes a timelike momentum. The approach of this paper extends naturally to spinning external states, where we expect multi-point positivity will lead to even more interesting bounds.

We will specifically focus on understanding the implications of positivity for the dimensionless version of the two-point energy correlator
\be
\boxed{F(z) \equiv \fr{\<\psi|\Ecal(\hat{n}_1) \; \Ecal(\hat{n}_2)|\psi\>}{\<\Ecal(\hat{n}_1)\> \; \<\Ecal(\hat{n}_2)\> \; \<\psi|\psi\>}\geq 0\;,}
\label{eq:FDef}
\ee
where we have normalized by the one-point function
\be
\<\Ecal(\hat{n})\> = \fr{\G(\fr{d-1}{2})}{2\pi^{\fr{d-1}{2}}} \; \fr{p^{d}}{(p^0 - \vec{p} \cdot \hat{n})^{d-1}}\;,
\ee
which is fixed by energy conservation. The two-point energy correlator is a function of the dimensionless cross-ratio
\be
z \equiv \fr{p^2(1 - \hat{n}_1\cdot \hat{n}_2)}{2(p^0 - \vec{p} \cdot \hat{n}_1) (p^0 - \vec{p} \cdot \hat{n}_2)}\,.
\label{eq:CrossRatio}
\ee
In the rest frame of the external source, $z = \half\big(1-\cos\theta\big)$, with $\theta$ the angle between the two detectors. 

This two-point function can be decomposed in terms of $SO(d-1)$ representations via the partial wave (or multipole) expansion,
\be
\boxed{F(z) = \sum_{\ell=0}^\infty F_\ell \, \Chat_\ell(1-2z),} \label{eq:PartialWaveExpansion}
\ee
where $\Chat_\ell(x)$ is the ($d$-dependent) Gegenbauer polynomial
\be
\Chat_\ell(x) \equiv \bigg(\fr{2\ell+d-3}{d-3}\bigg) \; C^{(\fr{d-3}{2})}_\ell(x).
\label{eq:Cdef}
\ee
The individual partial wave coefficients $F_\ell$ can be extracted from the energy correlator via the integral
\be
\boxed{F_\ell = \fr{1}{N_\ell} \int_0^1 dz \; \big(z(1-z)\big)^{\fr{d}{2}-2} \; F(z) \; \Chat_\ell(1-2z),}
\label{eq:FLdef}
\ee
where the normalization coefficient $N_\ell$ is given by
\be
N_\ell \equiv \int_0^1 dz \; \big(z(1-z)\big)^{\fr{d}{2}-2} \; \Big(\Chat_\ell(1-2z)\Big)^2 = \fr{\pi(2\ell+d-3) \; \G(\ell+d-3)}{\ell! \; 4^{d-3} \; \G^2(\fr{d-1}{2})}.
\label{eq:Ndef}
\ee
These partial wave coefficients were shown in~\cite{Dempsey:2025yiv} to be analytic in $\ell$, by constructing an inversion formula defining the $F_\ell$ in terms of an integral of the discontinuity of $F(z)$.

The energy correlator, and therefore its partial wave coefficients, depend on the quantum numbers of the chosen external state, including the total invariant mass $s \equiv p^2$. For different values of $s$, the values of the coefficients $F_\ell(s)$ generically will vary, tracing out a theory-dependent trajectory associated with the renormalization group flow of the QFT. The endpoints of these trajectories therefore correspond to RG fixed points, which are typically described by conformal field theories.

The goal of this work is to determine the allowed space for all possible trajectories corresponding to physical QFTs. In particular, we'd like to determine the allowed values for the partial wave coefficients $F_\ell$ given the requirement that the full two-point energy correlator must be non-negative. To start, we'll first review the standard bounds on integrals of positive functions against monomials, then extend this analysis to integrals against Gegenbauer polynomials. Finally, we'll use this procedure to derive universal bounds on energy correlator partial wave coefficients in all QFTs and the more specialized case of CFTs.


\subsection{Recap of Moment Positivity Bounds}

Suppose we have a function $\mu(z)$ whose only known property is that it is non-negative for $z \in [0,1]$,
\be
\mu(z) \geq 0 \qquad (0 \leq z \leq 1).
\ee
We can then ask about the possible values for the moments $a_n$ of this function, defined as
\be
a_n \equiv \int_0^1 dz \; \mu(z) \; z^n \qquad (n=0,1,\ldots).
\label{eq:ANdef}
\ee
Naively, the only bound we can place is that all $a_n$ must be non-negative, as we can rescale $\mu(z)$ (and therefore all $a_n$) by an arbitrary positive number. However, if we choose to normalize $\mu(z)$ such that
\be
a_0 \equiv \int_0^1 dz \; \mu(z) = 1,
\label{eq:MeasureNorm}
\ee
then we immediately find that all other $a_n$ must satisfy the bounds
\be
0 \leq a_n \leq 1,
\ee
due to the fact that $0 \leq z^n \leq 1$ in the range of interest.

We can actually further refine these bounds to determine the particular combinations of $a_n$ that are allowed by positivity of $\mu(z)$. Such bounds have been discussed extensively in the literature, in particular to constrain Wilson coefficients in weakly-coupled effective field theories~\cite{Bellazzini:2020cot,Tolley:2020gtv,Caron-Huot:2020cmc,Arkani-Hamed:2020blm}, so we will just briefly summarize the procedure for obtaining the optimal positivity bounds on combinations of $a_n$.

\begin{figure}[t!]
\centering
\includegraphics[width=.9\linewidth]{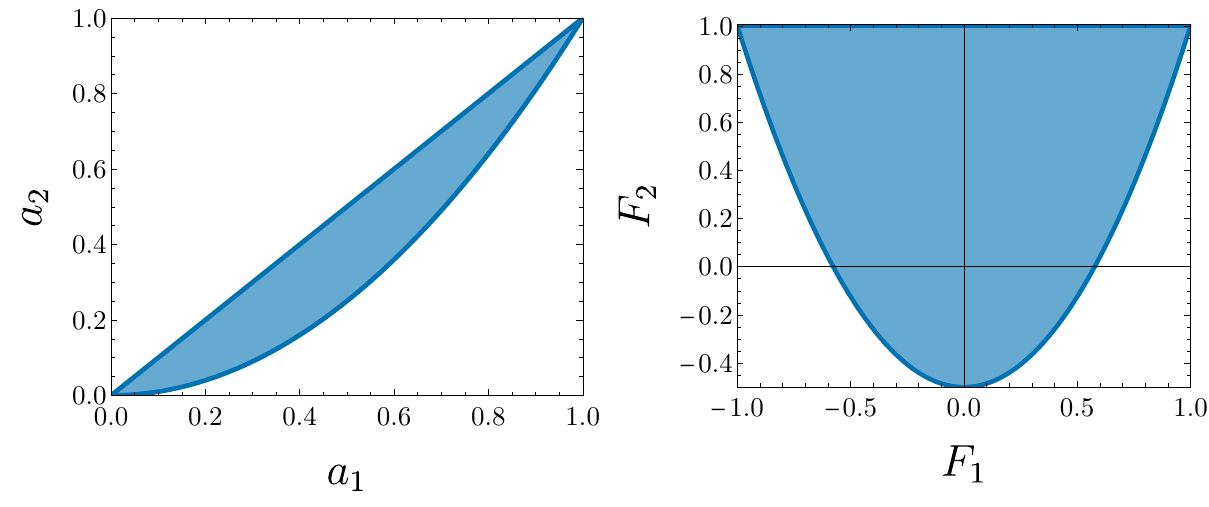}
\caption{\emph{Left:} Allowed region for the moments $a_1$ and $a_2$ obtained from a non-negative function $\mu(z)$. \emph{Right:} Allowed region for the $d=4$ partial wave coefficients (i.e.~Legendre polynomial coefficients) $F_1$ and $F_2$ obtained from a non-negative function $\mu(z)$. In both plots $a_0 = F_0 = 1$, and there is a one-to-one map between points in the allowed region of the left plot and those in the right plot.}
\label{fig:PositivityExample}
\end{figure}

First, we need to define the Hankel matrix $H^{\nmin}_{\nmax}$, with entries
\be
\big(H^{\nmin}_{\nmax}\big)_{ij} \equiv a_{\nmin+i+j} \qquad \big(i,j=0,1,\ldots,\lfloor \tfr{\nmax-\nmin}{2} \rfloor\big).
\label{eq:HankelDef}
\ee
For example, if we choose $\nmin=0$ and $\nmax=2$, we have
\be
H^0_2 = \begin{pmatrix} a_0 & a_1 \\ a_1 & a_2 \end{pmatrix}.
\label{eq:HankelExample}
\ee
To obtain the optimal bounds on all $a_n$ with $n \leq \nmax$, we need to demand that the following four matrices are positive semi-definite,
\be
\boxed{H^0_{\nmax} \succeq 0, \quad H^1_{\nmax} \succeq 0, \quad H^0_{\nmax-1} - H^1_{\nmax} \succeq 0, \quad H^1_{\nmax-1} - H^2_{\nmax} \succeq 0.}
\label{eq:PositivityBounds}
\ee
As a concrete example, the resulting bounds on $a_1$ and $a_2$ are
\be
a_1^2 \leq a_2 \leq a_1,
\ee
as shown in the left plot in Figure~\ref{fig:PositivityExample}. All positive measures $\mu(z)$ normalized according to~\eqref{eq:MeasureNorm} will only yield moments located in the allowed region shown in this plot.

\subsection{Extension to Partial Wave Coefficients}

We'd now like to apply these moment bounds to the case where the measure is built from the energy correlator,
\be
\mu(z) = \fr{1}{N_0} \; \big(z(1-z)\big)^{\fr{d}{2}-2} \; F(z).
\ee
However, in this case the natural observables aren't the moments $a_n$ but rather the partial wave coefficients
\be
F_\ell = \int_0^1 dz \; \mu(z) \; \fr{N_0}{N_\ell} \; \Chat_\ell(1-2z).
\ee
We therefore need to rewrite the entries of the Hankel matrices in~\eqref{eq:PositivityBounds} in terms of partial wave coefficients instead. We can do this by expanding each monomial in terms of Gegenbauer polynomials,
\be
z^n = \sum_{\ell=0}^n \zeta^n_\ell \; \fr{N_0}{N_\ell} \; \Chat_\ell(1-2z),
\label{eq:MonoToPartial}
\ee
where the coefficients are given by\footnote{In~\eqref{eq:Cdef}, \eqref{eq:Ndef}, and \eqref{eq:ZetaDef} the limit $\ell \ra 0$ should always be taken before $d \ra 3$, such that
\be
\Chat_0(1-2z) = 1, \quad N_0 = \fr{\pi \; \G(d-2)}{4^{d-3} \; \G^2(\fr{d-1}{2})}, \quad \zeta^n_0 = \fr{(\fr{d-2}{2})_n}{(d-2)_n},
\ee
for any $d\geq3$.}
\be
\zeta^n_\ell = (-1)^\ell \; \binom{n}{\ell} \; \fr{(2\ell+d-3) \; \G(\ell+d-3) \; \G(n+\fr{d-2}{2})}{\G(n+\ell+d-2) \; \G(\fr{d-2}{2})}.
\label{eq:ZetaDef}
\ee
Inserting~\eqref{eq:MonoToPartial} into~\eqref{eq:ANdef}, we therefore obtain the map from moments to partial wave coefficients,
\be
a_n = \sum_{\ell=0}^n \zeta^n_\ell \; F_\ell.
\ee
For example, in $d=4$ the Gegenbauer polynomials are simply Legendre polynomials,
\be
\Chat_\ell(1-2z) = (2\ell+1) P_\ell(1-2z) \qquad (d=4),
\ee
and the Hankel matrix $H^0_2$ from~\eqref{eq:HankelExample} can be rewritten as
\be
H^0_2 = \begin{pmatrix} F_0 & \tfr{1}{2}(F_0-F_1) \\ \tfr{1}{2}(F_0-F_1) & \tfr{1}{6}(2F_0-3F_1+F_2) \end{pmatrix}.
\ee
Unsurprisingly, when constructing the Hankel matrices the maximum moment power $\nmax$ corresponds to the maximum polynomial degree $\Lmax$.

The resulting positivity bounds on partial wave coefficients are obtained by demanding that the four matrices in~\eqref{eq:PositivityBounds} are positive semi-definite, but with the entries written in terms of $F_\ell$ instead of $a_n$. As an example, the resulting bounds on $F_1$ and $F_2$ in $d=4$ are
\be
\half(3F_1^2 - 1) \leq F_2 \leq 1,
\ee
as shown in right plot in Figure~\ref{fig:PositivityExample}.

\begin{figure}[t!]
\centering
\includegraphics[width=.9\linewidth]{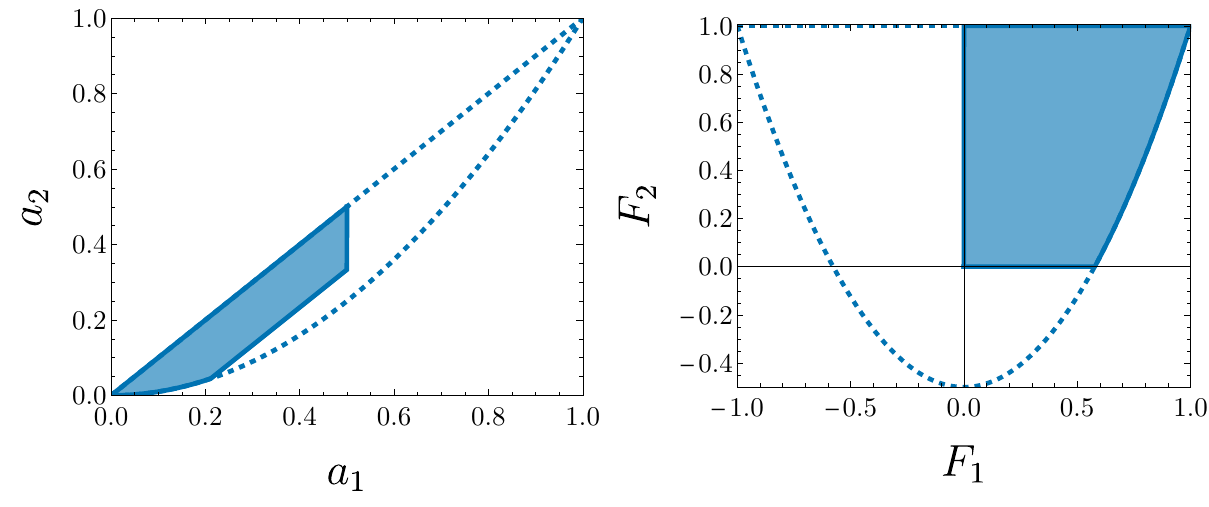}
\caption{Modification to the allowed regions in Figure~\ref{fig:PositivityExample} due to the additional constraints $F_1\geq0$ and $F_2\geq0$.}
\label{fig:UnitarityExample}
\end{figure}

However, while these are the optimal bounds placed solely by positivity of the energy correlator, there's an additional constraint that follows from unitarity. Specifically, every partial wave coefficient must be non-negative~\cite{Fox:1978vw},
\be
\boxed{F_\ell \geq 0.}
\label{eq:UnitarityBounds}
\ee
This is an \emph{independent} constraint that does \emph{not} automatically follow from energy positivity, as we can see in Figure~\ref{fig:UnitarityExample}, which shows the same example as Figure~\ref{fig:PositivityExample} but with the additional constraint that $F_1$ and $F_2$ are non-negative. To determine the allowed space of partial wave coefficients for energy correlators in unitary QFTs, we must therefore impose \emph{both}~\eqref{eq:PositivityBounds} and \eqref{eq:UnitarityBounds}.


\subsection{Positivity Bounds for QFT Energy Correlators}

We now have all the ingredients needed to determine the allowed space of partial wave coefficients for energy correlators in any unitary QFT. First, the value of the $\ell=0$ coefficient is fixed by conservation of energy, as integrating the energy flow operator over the celestial sphere simply measures the total energy of the external state. We have therefore normalized~\eqref{eq:FLdef} such that
\be
\boxed{F_0 = 1,}
\label{eq:EnergyCons}
\ee
for any external state.

We can now ask about the optimal bounds for the remaining coefficients. As a concrete example, let's consider the coefficients $F_1$ and $F_2$ for QFTs in $d=4$. In the previous sections we found that energy positivity alone restricted these coefficients to the region shown in Figure~\ref{fig:PositivityExample} and unitarity required $F_1$ and $F_2$ to be non-negative, reducing the allowed region to the one shown in Figure~\ref{fig:UnitarityExample}.

\begin{figure}[t!]
\centering
\includegraphics[width=.9\linewidth]{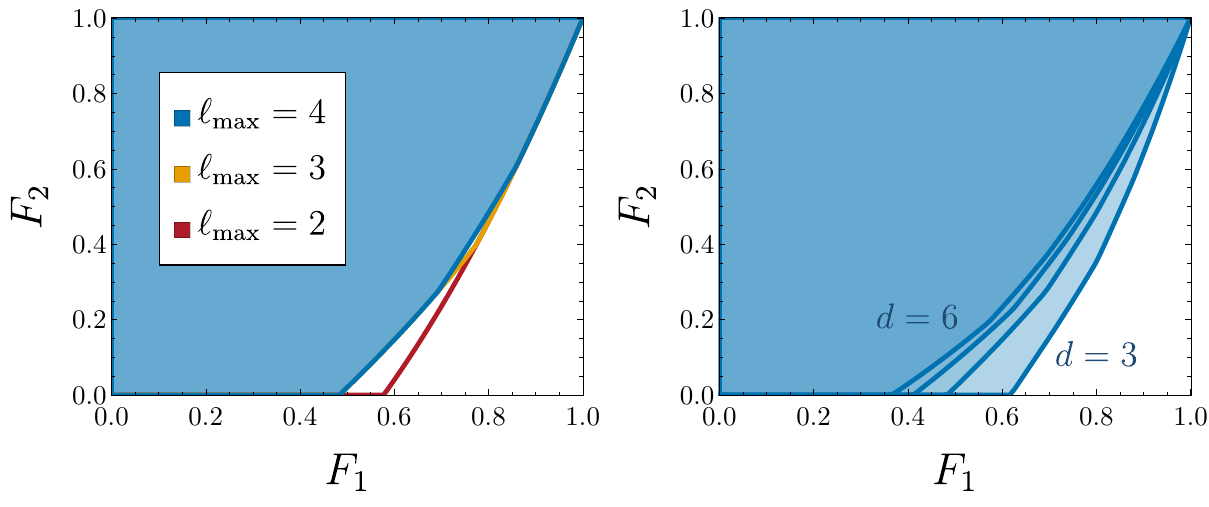}
\caption{\emph{Left:} Allowed region for the energy correlator partial wave coefficients $F_1$ and $F_2$ in any unitary QFT in $d=4$, computed up to $\Lmax=2$ (red), $\Lmax=3$ (yellow), and $\Lmax=4$ (blue). \emph{Right:} Allowed region for $F_1$ and $F_2$ in any unitary QFT in $d=3$, $4$, $5$, and $6$, where $d=3$ corresponds to the outermost bound and $d=6$ corresponds to the innermost bound, all computed with $\Lmax=10$.}
\label{fig:F1F2}
\end{figure}

It turns out, however, that those are \emph{not} the optimal bounds for these two coefficients. Instead, we can further improve the bounds by including $F_\ell$ with $\ell \geq 3$, requiring both the energy positivity restrictions in~\eqref{eq:PositivityBounds} and the non-negativity of all coefficients in~\eqref{eq:UnitarityBounds}. The left plot in Figure~\ref{fig:F1F2} shows the bounds on $F_1$ and $F_2$ obtained by including all coefficients $F_\ell$ with $\ell \leq \Lmax$, for different values of $\Lmax$. While in principle we should take $\Lmax \ra \infty$ to obtain the optimal bounds, in practice the effects of increasing $\Lmax$ quickly become negligible, as we can already see by comparing the $\Lmax=3$ and $\Lmax=4$ bounds in Figure~\ref{fig:F1F2} (see appendix~\ref{app:Lmax} for more details).

The right plot in Figure~\ref{fig:F1F2} shows the bounds on $F_1$ and $F_2$ for QFTs in different spacetime dimension $d$, all computed with $\Lmax=10$. The bounds become strictly stronger as we increase $d$, such that $d=3$ corresponds to the outermost bound and $d=6$ corresponds to the innermost bound. It is somewhat intuitive that increasing $d$ strictly reduces the allowed region, as QFTs in higher $d$ can be dimensionally reduced and reinterpreted as a theory in lower $d$ with an infinite number of Kaluza-Klein modes.

\begin{figure}[t!]
\centering
\includegraphics[width=.9\linewidth]{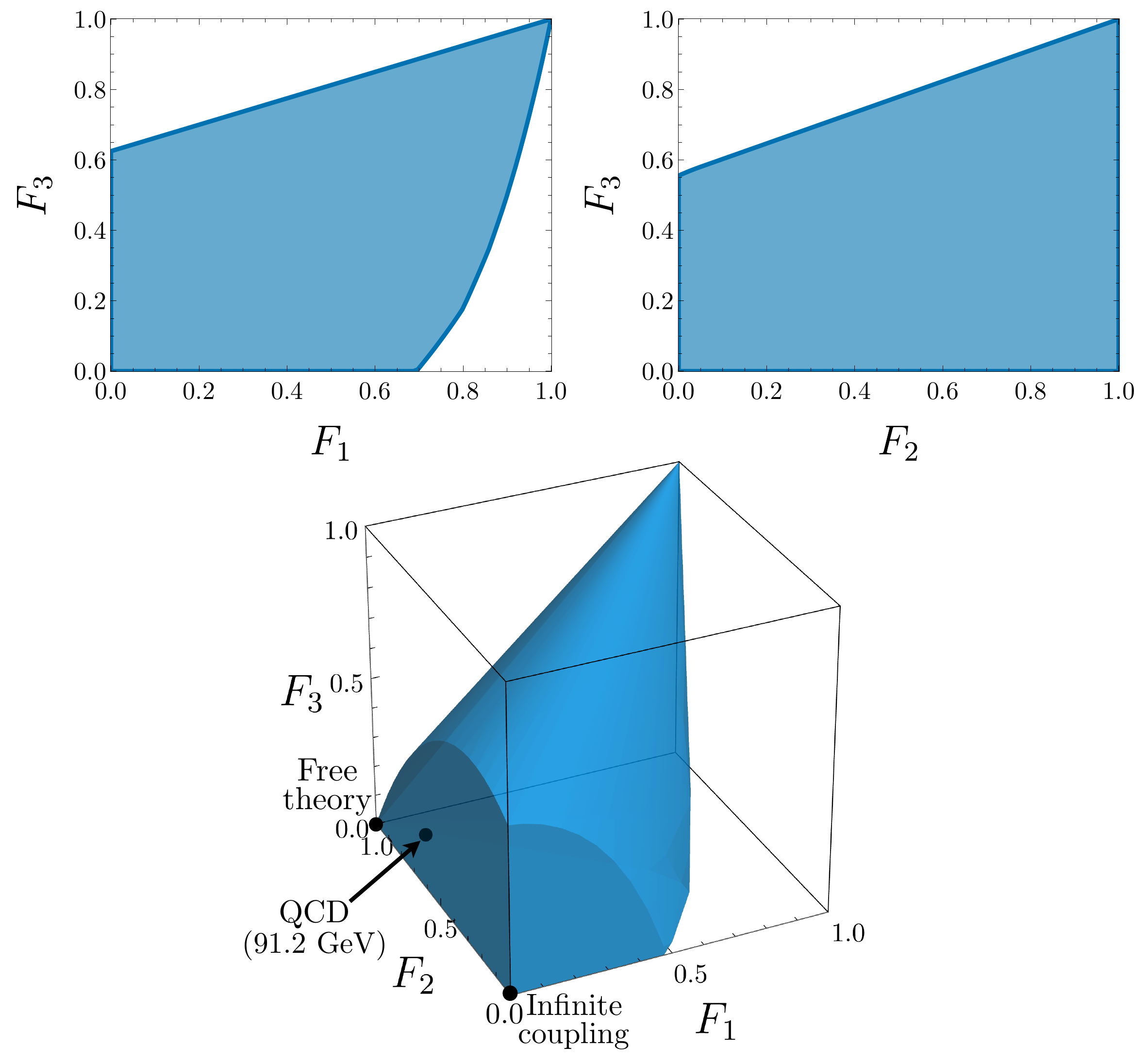}
\caption{\emph{Top:} Allowed region for the energy correlator partial wave coefficients $F_1$ and $F_3$ (\emph{left}) and $F_2$ and $F_3$ (\emph{right}) in any unitary QFT in $d=4$. \emph{Bottom:} Allowed region for $F_1$, $F_2$, and $F_3$ in any unitary QFT in $d=4$, highlighting the specific points in this region corresponding to a free theory, the infinite-coupling limit of planar $\Ncal=4$ super-Yang-Mills, and QCD at the $Z$-pole~\cite{Jaarsma:2025tck}. All bounds were computed with $\Lmax=10$.}
\label{fig:F1F2F3}
\end{figure}

We can place similar bounds on other combinations of partial wave coefficients, as well. The top row of Figure~\ref{fig:F1F2F3} shows the resulting bounds on $F_3$ as a function of both $F_1$ and $F_2$, all computed in $d=4$ with $\Lmax=10$. Of course, each of these bounds is just a two-dimensional projection of an infinite-dimensional region, which we call the ``EEC-hedron''. The bottom of Figure~\ref{fig:F1F2F3} shows the resulting three-dimensional projection of this EEC-hedron for $F_1$, $F_2$, and $F_3$. Any energy correlator in the background of any state in a unitary QFT corresponds to a point somewhere within this polyhedron.

For reference, we have highlighted three specific points in the allowed region shown in Figure~\ref{fig:F1F2F3}. The first point ($F_1=F_3=0$, $F_2=1$) corresponds to the energy correlator in the background of a two-particle state in free field theory, while the second point ($F_1=F_2=F_3=0$) corresponds to planar $\Ncal=4$ super-Yang-Mills at infinite coupling~\cite{Hofman:2008ar}. The third point corresponds to QCD at $\sqrt{s} = 91.2$ GeV, based on a recent analysis of archival LEP data~\cite{Electron-PositronAlliance:2025fhk}, with the resulting coefficients~\cite{Jaarsma:2025tck}
\be
F_1 = 0.056(5), \quad F_2 = 0.742(19), \quad F_3 = 0.092(5).
\ee
The radius of the marker for QCD has been chosen to match the largest uncertainty value ($0.019$), for visibility. As we vary the center-of-mass energy $s$, the point for QCD will move, tracing out a trajectory within the EEC-hedron. Because QCD is asymptotically free, it will approach the free theory point for $s \ra \infty$, and move closer to the infinite coupling point for $s \ra \Lambda_{\textrm{QCD}}^2$. It would be very interesting to eventually map out the full trajectory of QCD through this allowed region, building on both theoretical calculations~\cite{Dixon:2018qgp,Dixon:2019uzg,Chen:2020vvp,Ebert:2020sfi,Chang:2025zib,Chang:2025kgq} and experimental measurements~\cite{Komiske:2022enw,Lee:2022uwt,CMS:2024mlf,ALICE:2024dfl,STAR:2025jut}.


\subsection{Positivity Bounds for CFT Energy Correlators}

The bounds we've computed so far rely solely on energy positivity and unitarity, and therefore apply to any QFT. If we specialize to the case of a CFT, we find the additional restriction that the $\ell=1$ coefficient must vanish~\cite{Kologlu:2019mfz,Korchemsky:2019nzm},
\be
\boxed{F_1 = 0,}
\label{eq:MomCons}
\ee
which is a consequence of momentum conservation for massless degrees of freedom.

\begin{figure}[t!]
\centering
\includegraphics[width=.9\linewidth]{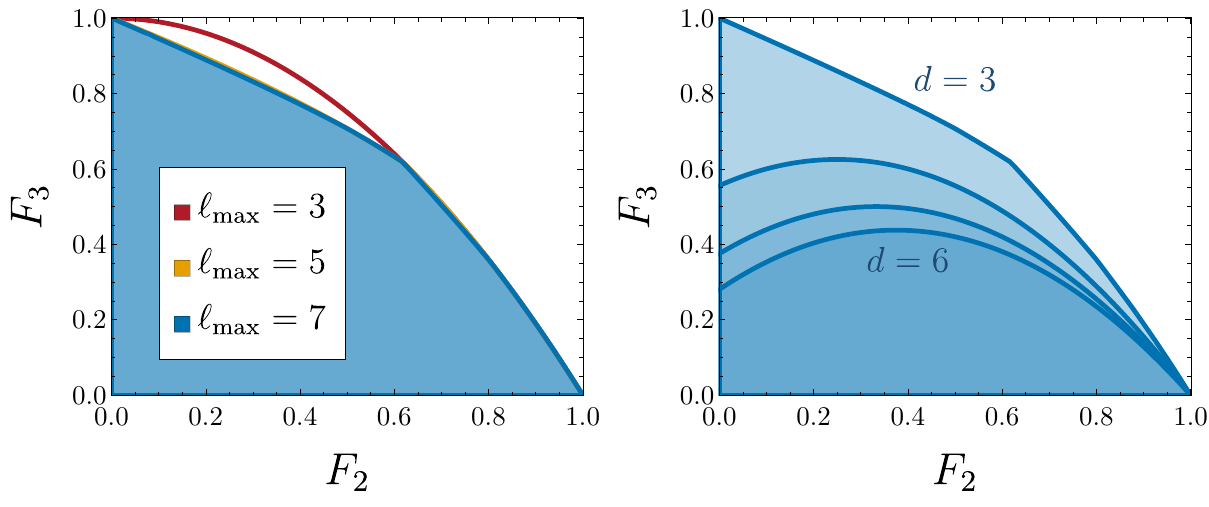}
\caption{\emph{Left:} Allowed region for the energy correlator partial wave coefficients $F_2$ and $F_3$ in any unitary CFT in $d=3$, computed up to $\Lmax=3$ (red), $\Lmax=5$ (yellow), and $\Lmax=7$ (blue). \emph{Right:} Allowed region for $F_2$ and $F_3$ in any unitary CFT in $d=3$, $4$, $5$, and $6$, where $d=3$ corresponds to the outermost bound and $d=6$ corresponds to the innermost bound, all computed with $\Lmax=10$.}
\label{fig:F2F3}
\end{figure}

We can derive bounds on the remaining partial wave coefficients by following the same procedure as for more general QFTs, with the addition of this new constraint. For example, the left plot in Figure~\ref{fig:F2F3} shows the bounds on $F_2$ and $F_3$ in 3d CFTs obtained by including all coefficients $F_\ell$ with $\ell \leq \Lmax$. As we can see, the bounds again improve as we increase $\Lmax$. From the right plot in Figure~\ref{fig:F2F3}, we also see that the bounds again become strictly stronger as the spacetime dimension $d$ is increased. Interestingly, for $d \geq 4$ we see that positivity places an absolute upper bound on $F_3$ that is stronger than the naive one of $F_3 \leq 1$,
\be
F_3 \leq \fr{d+1}{4(d-2)},
\ee
as was first observed in~\cite{Dempsey:2025yiv}. Of course, we could have also determined this directly from the bounds on $F_3$ at $F_1 = 0$ in the upper left plot in Figure~\ref{fig:F1F2F3}. Similar absolute upper bounds can be computed for all $F_\ell$ with odd values of $\ell$.

\begin{figure}[t!]
\centering
\includegraphics[width=.9\linewidth]{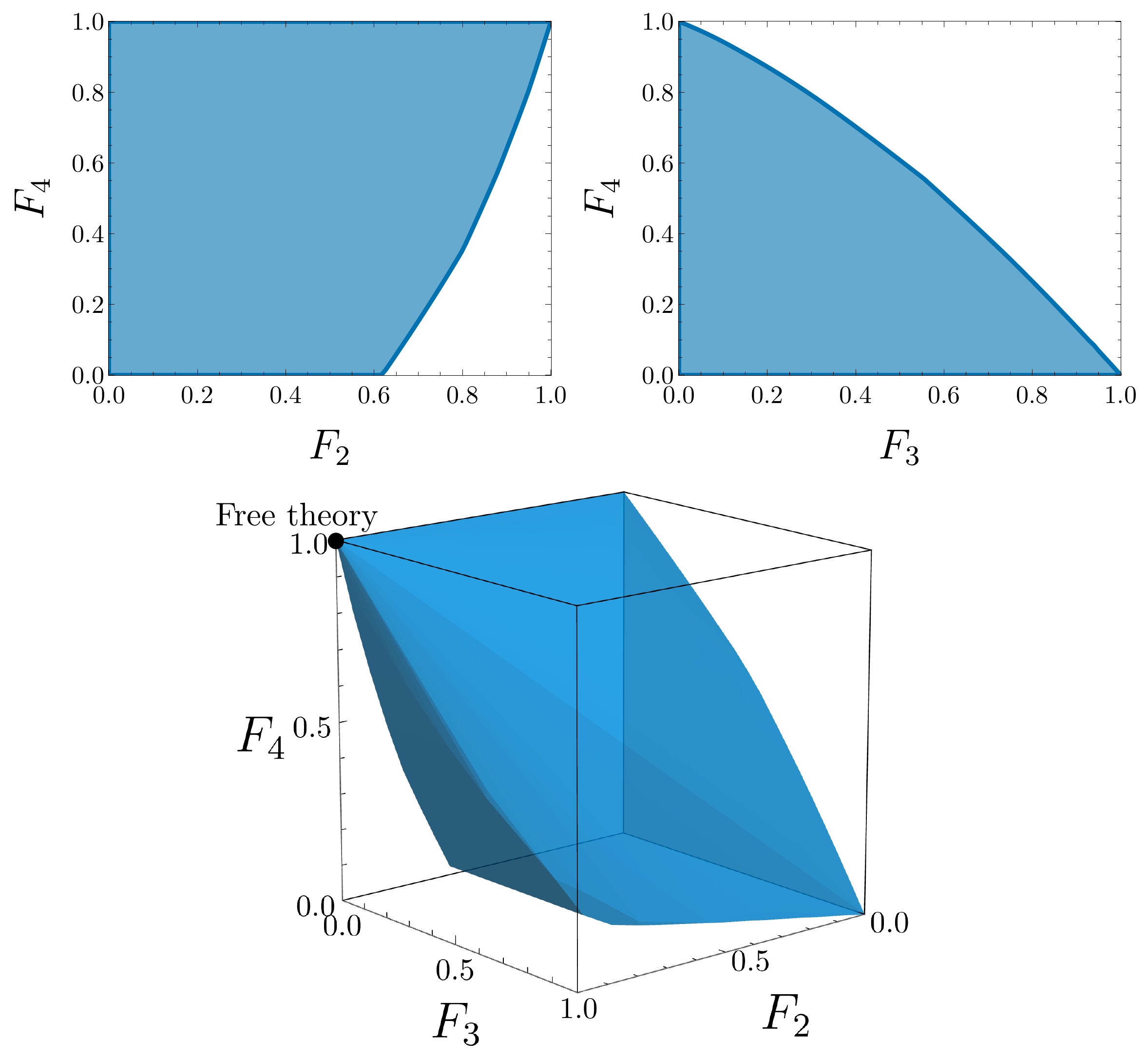}
\caption{\emph{Top:} Allowed region for the energy correlator partial wave coefficients $F_2$ and $F_4$ (\emph{left}) and $F_3$ and $F_4$ (\emph{right}) in any unitary CFT in $d=3$. \emph{Bottom:} Allowed region for $F_2$, $F_3$, and $F_4$ in any unitary CFT in $d=3$, highlighting the specific point in this region corresponding to a free CFT. All bounds were computed with $\Lmax=10$.}
\label{fig:F2F3F4}
\end{figure}

Figure~\ref{fig:F2F3F4} shows the bounds on the higher partial wave coefficient $F_4$ as a function of $F_2$ and $F_3$, both as two-dimensional projections (top row) and the full three-dimensional allowed region (bottom). For reference, we have highlighted the specific point ($F_2=F_4=1$, $F_3=0$) in this allowed region corresponding to the energy correlator in the background of a two-particle state in a free CFT. This procedure can in principle be applied to all higher values of $\ell$ in order to determine the full EEC-hedron for 3d CFTs.

Because CFTs are scale-invariant, the partial wave coefficients $F_\ell$ no longer depend on the invariant mass $s$ of the external state. The energy correlator for a given state at any energy therefore corresponds to a specific point within the allowed region shown in Figure~\ref{fig:F2F3}, rather than an entire trajectory as in more general QFTs.


\section{Bounds on OPE Coefficients in the 3d Ising CFT}
\label{sec:IsingBounds}

In the previous section, we demonstrated how energy positivity and unitarity place strong constraints on the partial wave coefficients of energy correlators in general QFTs. Now we turn our attention to a specific theory, the 3d Ising CFT, and consider the energy correlator in the background of the lowest-dimension $\mathbb{Z}_2$-even scalar operator $\vep$. By combining the EEC-hedron with data from the conformal bootstrap~\cite{Simmons-Duffin:2016wlq,Chang:2024whx,Poland:2025ide} and fuzzy sphere regularization~\cite{Fardelli:2026zas}, we are able to obtain novel constraints on the OPE coefficients of operators in the $\vep \times T$ OPE.


\subsection{Recap of Conformal Block Expansion and Conventions}
\label{sec:Recap}

In a CFT, the energy correlator $F(z)$ can be decomposed into irreducible representations of the conformal group, each labeled by a primary operator $\Ocal$,
\be
F(z) = \sum_\Ocal \lambda_{\phi T\Ocal}^2 \; G_\Ocal(z),\label{eq:ConformalBlockExpansion}
\ee
where $\phi$ is the scalar primary creating the external state, as defined in~\eqref{eq:PhiDef}. The contribution of each irreducible representation (i.e.~$\Ocal$ and its descendants $\p^n\Ocal$) is given by a theory-independent function known as a \emph{conformal block} $G_\Ocal(z)$, which is completely fixed by symmetry, multiplied by a theory-dependent \emph{OPE coefficient} $\lambda_{\phi T\Ocal}$. Because we are specifically interested in the 3d Ising CFT, which is invariant under the parity transformation $\vec{x} \ra -\vec{x}$, each operator $\Ocal$ in this sum must be either even or odd under parity.

Each conformal block can itself be further decomposed into representations of $SO(d-1)$ (i.e.~partial waves), 
\be
G_\Ocal(z) = \sum_{\ell=2}^J G^{\Ocal}_\ell \; \Chat_\ell(1-2z), \label{eq:ConfBlockPWDec}
\ee
where $J\geq2$ is the spin of the primary operator $\Ocal$.\footnote{As discussed in~\cite{Mecaj:2025ecl}, due to Ward identities the only operator with $J < 2$ which contributes to the energy correlator $F(z)$ is $\phi$ itself, with the corresponding conformal block $G_\phi(z) = G^\phi_0$.} Each energy correlator partial wave coefficient $F_\ell$ with $\ell \geq 2$ therefore has its own conformal block decomposition
\be
\boxed{F_\ell = \sum_\Ocal \lambda_{\phi T\Ocal}^2 \; G^\Ocal_\ell,}
\label{eq:ConfBlockFL}
\ee
receiving contributions from all primary operators with spin $J \geq \ell$.

The conformal block coefficients $G^\Ocal_\ell$ for any traceless symmetric operator $\Ocal$ were recently computed in~\cite{Mecaj:2025ecl}, building on the general procedure developed in~\cite{Kologlu:2019bco}. In this work we'll be focused on the case where $\phi$ is parity-even, with the following convention for OPE coefficients $\lambda_{\phi T\Ocal^+}$ with other parity-even operators $\Ocal^+$,
\be
\<\phi(x_1) \; T(x_2) \; \Ocal^+(x_3)\> = \lambda_{\phi T\Ocal^+} \Bigg(\fr{V_{2,13}^2 V_{3,12}^J + c_1^+ \; V_{2,13} V_{3,12}^{J-1} H_{23} + c_2^+ \; V_{3,12}^{J-2} H_{23}^2}{x_{12}^{\De_\phi-\De-J+d+2} x_{23}^{\De+J-\De_\phi+d+2} x_{13}^{\De_\phi+\De+J-d-2}}\Bigg).
\label{eq:OPEDefEven}
\ee
The coefficients $c_1$ and $c_2$ in this expression are fixed by Ward identities to
\be
\ba
c_1^+ &= -\fr{2\big((d-1)(\De-\De_\phi)+J\big)}{(d-2)(\De-J-\De_\phi-d)}, \\
c_2^+ &= \fr{(d-1)(\De-\De_\phi)^2-(d-2)J-J^2}{(d-2)(\De-J-\De_\phi-d)(\De-J-\De_\phi-d+2)},
\ea
\ee
and $V_{i,jk}$ and $H_{ij}$ are the standard tensor structure building blocks defined in~\cite{Costa:2011mg}. In $d=3$, there can also be non-zero OPE coefficients with parity-odd operators $\Ocal^-$, which in this work we define according the convention,
\be
\<\phi(x_1) \; T(x_2) \; \Ocal^-(x_3)\> = \lambda_{\phi T\Ocal^-} \Bigg(\fr{V_{2,13} V_{3,12}^{J-1} S_{23,1} + c_1^- \; V_{3,12}^{J-2} H_{23} \; S_{23,1}}{x_{12}^{\De_\phi-\De-J+5} x_{23}^{\De+J-\De_\phi+5} x_{13}^{\De_\phi+\De+J-5}}\Bigg),
\label{eq:OPEDefOdd}
\ee
where the coefficient $c_-$ is again fixed by Ward identities,
\be
c_1^- = - \fr{\De-\De_\phi}{\De-J-\De_\phi-2},
\ee
and the parity-odd tensor structure $S_{ij,k}$ is defined in~\cite{Mecaj:2025ecl} (see also~\cite{Costa:2011mg,Giombi:2011rz}).

In appendix~\ref{app:OddBlocks} we show that the conformal block coefficients for parity-odd operators are actually \emph{proportional} to those for parity-even operators,
\be
G_\ell^{\Ocal^-} = \fr{\G^2\big(\fr{\De_\phi-\De+J+5}{2}\big) \; \G^2\big(\fr{\De+J-\De_\phi+1}{2}\big)}{4\G^2\big(\fr{\De_\phi-\De+J+4}{2}\big) \; \G^2\big(\fr{\De+J-\De_\phi+2}{2}\big)} \; G_\ell^{\Ocal^+},
\label{eq:OddEvenRelation}
\ee
which was not manifest in the original derivation presented in~\cite{Mecaj:2025ecl}.

The most important property of these conformal block coefficients $G^\Ocal_\ell$ is that they are all non-negative,
\be
\boxed{G^\Ocal_\ell \geq 0,}
\ee
which means that the EEC-hedron bounds on $F_\ell$ derived in section~\ref{sec:EECBounds} automatically place \emph{upper bounds} on the values of OPE coefficients. As we'll see, these bounds can be significantly improved with some finite set of CFT data, which can be obtained via other methods such as the conformal bootstrap and fuzzy sphere regularization, making energy correlators a powerful and complementary means of constraining the allowed space of CFTs.

Let us make one final comment regarding our conventions for the normalization of the stress tensor $T$. The conformal block coefficients $G^\Ocal_\ell$ computed in~\cite{Mecaj:2025ecl}, as well as the associated OPE coefficients $\lambda_{\phi T\Ocal}$ in~\eqref{eq:OPEDefEven} and~\eqref{eq:OPEDefOdd}, have been normalized such that the stress tensor measures the total energy. As a result, the stress tensor two-point function is
\be
\<T(x) \; T(0)\> = c_T \; \fr{H^2}{x^{2(d+2)}},
\ee
where $H$ is again the standard tensor structure from~\cite{Costa:2011mg} and $c_T$ is the central charge of the CFT, which in the specific case of the 3d Ising CFT is~\cite{Chang:2024whx}
\be
c_T = 0.00899103927(40).
\ee
In order to directly compare our results to others in the bootstrap literature, we will define the ``unit-normalized'' stress tensor as
\be
\hat{T} \equiv \fr{T}{\sqrt{c_T}},
\label{eq:THatDef}
\ee
with the associated OPE coefficients
\be
\boxed{\lambda_{\phi\hat{T}\Ocal} \equiv \fr{\lambda_{\phi T\Ocal}}{\sqrt{c_T}}.}
\label{eq:LambdaHatDef}
\ee
For the remainder of this section, all reported values and bounds for OPE coefficients will specifically refer to $\lambda_{\phi\hat{T}\Ocal}$, but these can be converted to $\lambda_{\phi T\Ocal}$ via~\eqref{eq:LambdaHatDef}.


\subsection{Improved Bounds on Partial Waves from Bootstrap Data}
\label{sec:IsingPartialWave}

\begin{table}[t!]
\begin{center}
\begin{tabular}{ |l|l|l|l||l|l| } 
\hline
& & & & & \\[-13pt]
$\Ocal$ & $\De$ & $J$ & $P$ & $|\lambda_{\vep\hat{T}\Ocal}|$ & Upper bound \\[1pt] 
\hline
$\vep$ & $1.41262528(29)$~\cite{Chang:2024whx} & $0$ & $+$ & $1.7782942(80)$~\cite{Chang:2024whx} & \\
\hline
& & & & & \\[-13pt]
$\hat{T}$ & $3$ & $2$ & $+$ & $0.95331513(42)$~\cite{Chang:2024whx} & 
\\
$T'$ & $5.50915(44)$~\cite{Simmons-Duffin:2016wlq} & $2$ & $+$ & $0.0300(44)$~\cite{Fardelli:2026zas} & 
$0.252$ \\
$T''$ & $7.0758(58)$~\cite{Simmons-Duffin:2016wlq} & $2$ & $+$ & $0.0049(35)$~\cite{Fardelli:2026zas} & 
$0.538$ \\
$T^-$ & $7.41(18)$~\cite{Fardelli:2026zas} & $2$ & $-$ & & \\
\hline
$C$ & $5.022665(28)$~\cite{Simmons-Duffin:2016wlq} & $4$ & $+$ & $0.459(15)$~\cite{Poland:2025ide} & 
$0.489$ \\
$C'$ & $6.42065(64)$~\cite{Simmons-Duffin:2016wlq} & $4$ & $+$ & $0.00077(30)$~\cite{Fardelli:2026zas} & 
$0.349$ \\
$C''$ & $7.38568(28)$~\cite{Simmons-Duffin:2016wlq} & $4$ & $+$ & $0.015(4)$~\cite{Fardelli:2026zas} & 
$0.309$ \\
$C^-$ & $8.04(2)$~\cite{Fardelli:2026zas} & $4$ & $-$ &  & 
$2.299$ \\
$C'''$ & $8.9410(99)$~\cite{Simmons-Duffin:2016wlq} & $4$ & $+$ &  & 
$0.522$ \\
\hline
$\vep_6$ & $7.028488(16)$~\cite{Simmons-Duffin:2016wlq} & $6$ & $+$ & $0.187(8)$~\cite{Poland:2025ide} & 
$0.243$ \\
$\vep_6'$ & $8.4957(75)$~\cite{Simmons-Duffin:2016wlq} & $6$ & $+$ &  & 
$0.233$ \\
$\vep_6''$ & $9.32032(34)$~\cite{Simmons-Duffin:2016wlq} & $6$ & $+$ & & 
$0.254$ \\
\hline
$\vep_8$ & $9.031023(30)$~\cite{Simmons-Duffin:2016wlq} & $8$ & $+$ &  & 
 $0.121$ \\
\hline
\end{tabular}
\end{center}
\caption{Known primary operators $\Ocal$ with $\De < 10$ in the $\vep \times T$ OPE, with the associated scaling dimension $\De$, spin $J$, parity $P$, the OPE coefficient magnitude $|\lambda_{\vep \hat{T}\Ocal}|$ (if known),\protect\footnotemark~and the upper bound on $|\lambda_{\vep \hat{T}\Ocal}|$ computed using the method outlined in section \ref{sec:IsingUpperBounds}. Note that $\hat{T} \equiv \fr{1}{\sqrt{c_T}} T$, as defined in~\eqref{eq:THatDef}.}
\label{tab:IsingData}
\end{table}

Before bounding individual OPE coefficients, we can first determine how 3d Ising CFT data previously obtained with the conformal bootstrap constrains the partial wave coefficients $F_\ell$. Concretely, let's consider the case where the external operator $\phi$ is the $\mathbb{Z}_2$-even scalar primary with the lowest scaling dimension: $\vep$.

\footnotetext{The values and uncertainties for OPE coefficients computed with the conformal bootstrap~\cite{Chang:2024whx,Poland:2025ide} are taken directly from the cited results, while the listed values for OPE coefficients obtained with fuzzy sphere regularization~\cite{Fardelli:2026zas} correspond to an extrapolation in the number of fermions $N \ra \infty$, with the uncertainty given by the difference between the extrapolation and the result with the largest value of $N$.\label{foot:ErrorBars}}

Table~\ref{tab:IsingData} shows the known low-dimension primary operators $\Ocal$ in the $\vep \times T$ OPE, along with their scaling dimension, spin $J$, parity $P$, and OPE coefficient $\lambda_{\vep\hat{T}\Ocal}$ (if known), normalized according to~\eqref{eq:LambdaHatDef}. From~\eqref{eq:ConfBlockFL}, we know that each of these operators provides a positive contribution to various partial wave coefficients $F_\ell$, so we can use this CFT data to bound the energy correlator in the background of $\vep$.

\begin{figure}[t!]
\centering
\includegraphics[width=.9\linewidth]{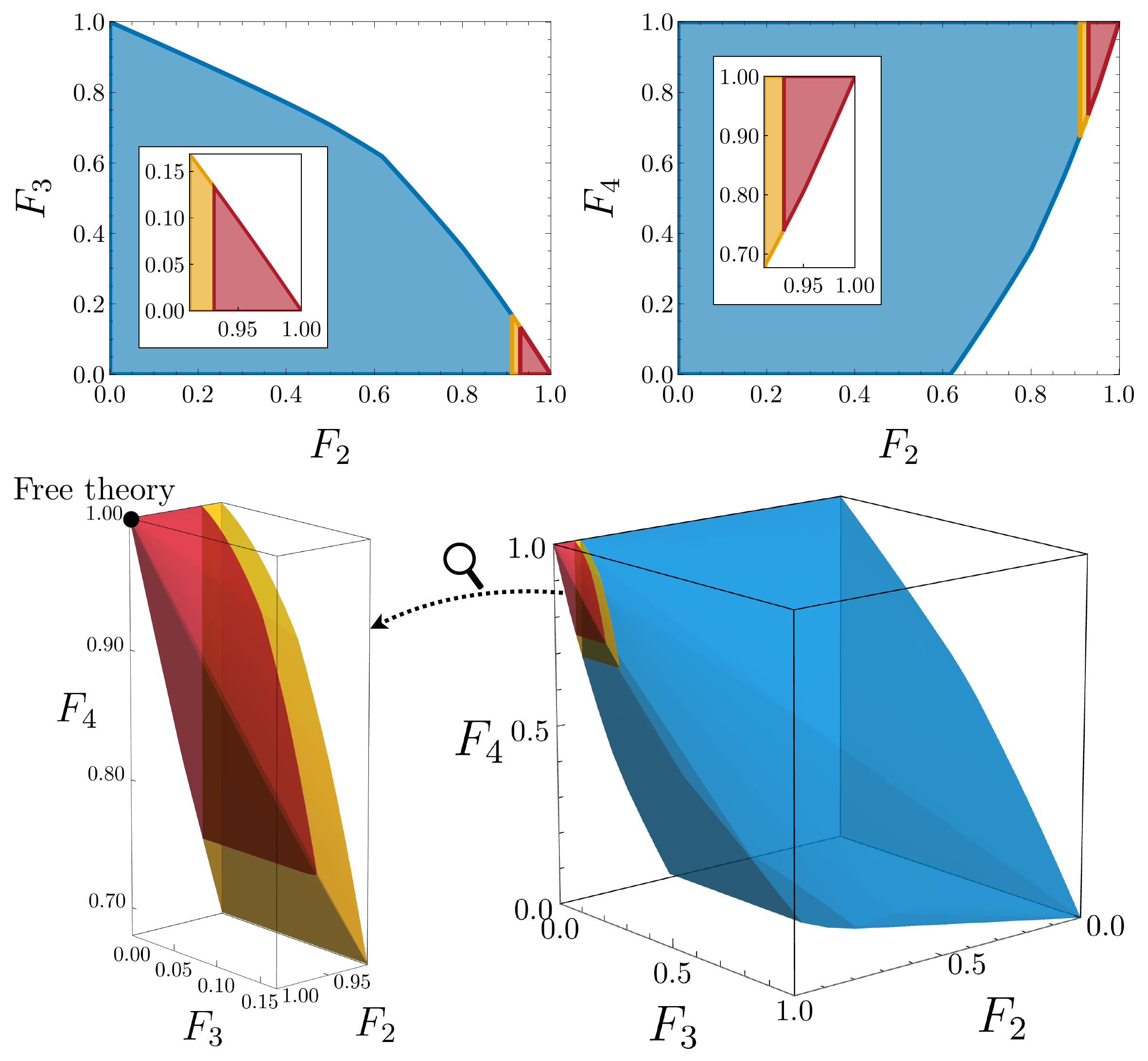}
\caption{\emph{Top:} Allowed region for the energy correlator partial wave coefficients $F_2$ and $F_3$ (\emph{left}) and $F_2$ and $F_4$ (\emph{right}) in the 3d Ising CFT (red) and the 3d $O(2)$ CFT (red and yellow), compared with the bounds for arbitrary 3d CFTs from Figures~\ref{fig:F2F3} and \ref{fig:F2F3F4} (blue). \emph{Bottom:} Allowed region for $F_2$, $F_3$, and $F_4$ in the 3d Ising CFT (red) and the 3d $O(2)$ CFT (red and yellow) compared with the bound for all 3d CFTs from Figure~\ref{fig:F2F3F4} (blue), highlighting for comparison the specific point in this allowed region corresponding to a free CFT. The insets show zoomed-in versions of the allowed region, and all bounds were computed with $\Lmax=10$.}
\label{fig:F2F3F4IsingO2}
\end{figure}

One important operator in this OPE is the stress tensor itself, whose OPE coefficient $\lambda_{\vep T T}$ was computed in~\cite{Chang:2024whx}, allowing us to place the following \emph{lower bound} on the $\ell=2$ coefficient~\cite{Mecaj:2025ecl},
\be
\boxed{F_2 \geq \lambda_{\vep TT}^2 \; G_2^T = 0.931,}
\ee
We can now combine this lower bound on $F_2$ with the positivity constraints from section~\ref{sec:EECBounds} to bound the remaining $F_\ell$. For example, the top left plot of Figure~\ref{fig:F2F3F4IsingO2} shows that this lower bound on $F_2$ leads to an \emph{upper bound} on $F_3$, significantly restricting the allowed region as compared to Figure~\ref{fig:F2F3}. Similarly, the top right plot of Figure~\ref{fig:F2F3F4IsingO2} shows that the lower bound on $F_2$ results in a \emph{lower bound} on $F_4$, again restricting the allowed region as compared to Figure~\ref{fig:F2F3F4}. This behavior continues for higher partial waves, resulting in new upper bounds for all odd $\ell$ and lower bounds for all even $\ell$, with the first few corresponding to:
\be
\boxed{F_3 \leq 0.132, \quad F_4 \geq 0.744, \quad F_5 \leq 0.360, \quad F_6 \geq 0.440, \quad F_7 \leq 0.624.}
\label{eq:FLBounds}
\ee

As we can see from the bottom plot of Figure~\ref{fig:F2F3F4IsingO2}, these new bounds restrict the energy correlator in the background of $\vep$ to lie in a small region, highlighted in red, compared to the full space of energy correlators. For comparison, we have indicated the specific point in this region ($F_2=F_4=1$, $F_3=0$) corresponding to the energy correlator in the background of $\phi^2$ in the free real scalar CFT. Even though the Ising CFT is strongly-coupled, it is interesting that its allowed values lie very close to that of a free theory.


\subsection{Upper Bounds on OPE Coefficients}
\label{sec:IsingUpperBounds}

We are now in a position to compute upper bounds on the OPE coefficients $\lambda_{\vep T\Ocal}$, using the fact that the conformal block partial wave coefficients $G_\ell^\Ocal$ from~\eqref{eq:ConfBlockPWDec} are manifestly positive~\cite{Mecaj:2025ecl}. From the conformal block decomposition of the energy correlator partial wave coefficient $F_\ell$ in~\eqref{eq:ConfBlockFL}, we find the following constraint relating any $F_\ell$ and an OPE coefficient $\lambda_{\vep T\Ocal}$:
\begin{equation}
    \lambda^2_{\vep T\Ocal} \; G_\ell^\Ocal \leq F_\ell. \label{eq:PositiveBlocks}
\end{equation}
The upper bounds on $F_\ell$ obtained from imposing positivity~\eqref{eq:PositivityBounds} and unitarity~\eqref{eq:UnitarityBounds} therefore automatically result in an upper bound on $|\lambda_{\vep T\Ocal}|$.

As discussed in section~\ref{sec:IsingPartialWave}, the OPE coefficient $\lambda_{\vep TT}$ has been computed to high precision with the conformal bootstrap~\cite{Chang:2024whx}, and we can now use this known value to strengthen our upper bound on $|\lambda_{\vep T\Ocal}|$ for all other operators in two ways. First, $\lambda_{\vep TT}$ provides a positive contribution to $F_2$, modifying the right-hand side of~\eqref{eq:PositiveBlocks} for $\ell=2$ to the improved constraint:
\be
\lambda_{\vep T\Ocal}^2 \; G_2^\Ocal \leq F_2 - \lambda_{\vep TT}^2 \; G_2^T.
\ee
Second, the contribution from the $T$ conformal block to $F_2$ leads to stronger upper bounds on all odd $F_\ell$, as listed in~\eqref{eq:FLBounds}, and therefore stronger bounds on all other OPE coefficients.

The rightmost column in Table~\ref{tab:IsingData} lists the resulting upper bounds on the OPE coefficients of all known operators with $\De<10$ in the $\vep \times T$ OPE. While for most operators, the uncertainty in the associated scaling dimension is small enough to have no effect on the upper bound's first few digits, this is not the case for $T^-$, the lowest-twist parity-odd operator with $J=2$. Interestingly, the scaling dimension $\De_{T^-}=7.41(18)$ computed in~\cite{Fardelli:2026zas} is remarkably close to the parity-odd ``double-twist'' scaling dimension $\De_\vep + 6 = 7.41262528(29)$,
where the conformal block coefficient $G_2^{\Ocal^-}$ vanishes~\cite{Mecaj:2025ecl}. As a result, no meaningful upper bound can be placed on this particular OPE coefficient. It would be useful to determine the value of $\De_{T^-}$ more precisely with the conformal bootstrap, in order to obtain an upper bound on its OPE coefficient.

\begin{figure}[!t]
\centering
\includegraphics[width=0.95\linewidth]{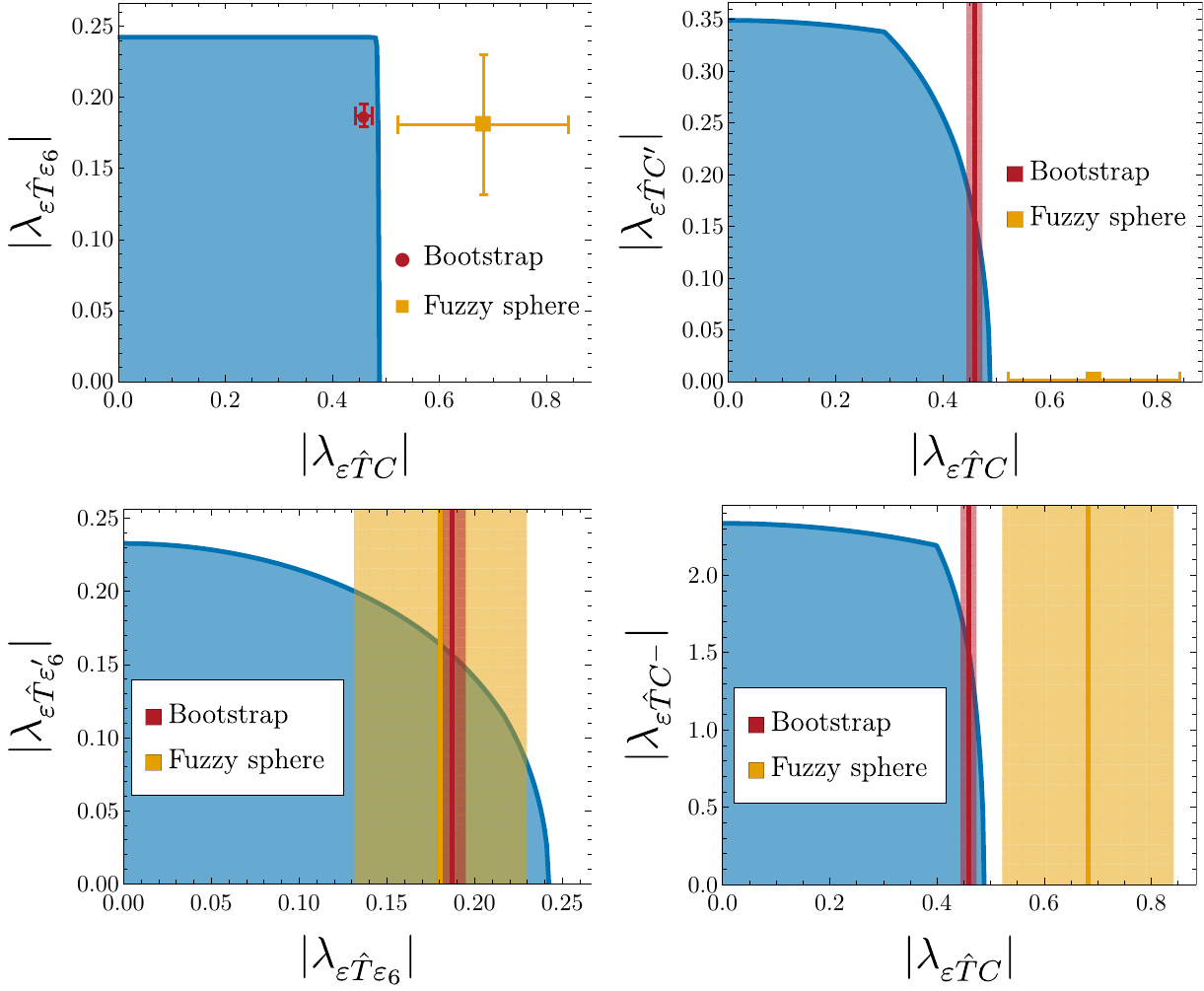}
\caption{Allowed values for the OPE coefficients of low-twist spin-$4$ operators ($C$, $C'$, $C^-$) and spin-$6$ operators ($\vep_6$, $\vep'_6$) in the 3d Ising CFT (blue), compared with known values (if available) computed with the conformal bootstrap~\cite{Poland:2025ide} (red) and fuzzy sphere regularization~\cite{Fardelli:2026zas} (yellow).}
\label{fig:IsingBounds}
\end{figure}

Of course, the upper bounds listed in Table~\ref{tab:IsingData} cannot all be saturated, as multiple operators contribute to the same $F_\ell$. We can instead obtain joint bounds on the OPE coefficients of two operators $\Ocal_1$ and $\Ocal_2$ by including both on the left-hand side of~\eqref{eq:PositiveBlocks}, resulting in the new constraint:
\be
\lambda^2_{\vep T\Ocal_1} G_\ell^{\Ocal_1} + \lambda^2_{\vep T\Ocal_2} G_\ell^{\Ocal_2} \leq F_\ell. \label{eq:JointBounds}
\ee
Figure~\ref{fig:IsingBounds} shows some examples of these joint bounds for known operators in the 3d Ising CFT, compared with the values (if known) computed with the conformal bootstrap~\cite{Poland:2025ide} and fuzzy sphere regularization~\cite{Fardelli:2026zas} (see footnote~\ref{foot:ErrorBars} regarding the determination of these values and uncertainties).

The upper left plot shows the joint bounds for $C$ and $\vep_6$, which are the lowest-twist operators with $J=4$ and $J=6$, respectively. As we can see, our upper bound on $|\lambda_{\vep TC}|$ actually \emph{excludes} the central value obtained with fuzzy sphere regularization, while the bootstrap value is quite close to saturating our bound. The upper right plot shows similar bounds for $C$ and $C'$, which are the two lowest-twist spin-$4$ operators. While there is currently no value for $\lambda_{\vep TC'}$ from the bootstrap, the value obtained from fuzzy sphere regularization is orders of magnitude smaller than the naive upper bound obtained from positivity, indicating that these bounds are most effective for the lowest-twist operator at each spin.

Similarly, the lower left plot shows the upper bounds for $\vep_6$ and $\vep'_6$, which are the two lowest-twist spin-$6$ operators. In this case, both the conformal bootstrap and fuzzy sphere regularization yield values for $\lambda_{\vep T\vep_6}$ that are comparable in size to the upper bound obtained from positivity, though there is no known value for $\lambda_{\vep T\vep'_6}$ from either method. Finally, the lower right plot shows the upper bounds for $C$ and $C^-$, which are the lowest-twist spin-$4$ operators with even and odd parity, respectively. These last two plots, in particular, show how energy positivity can be a useful complement to other methods for computing CFT data. Given a value for one OPE coefficient (e.g.~$\lambda_{\vep TC}$), we can immediately obtain an upper bound on other unknown OPE coefficients (e.g.~$\lambda_{\vep TC^-}$).


\subsection{Lower Bounds on OPE Coefficients}
\label{sec:IsingLowerBounds}

Naively, positivity of energy correlators appears to only place upper bounds on individual OPE coefficients, since the partial wave coefficients $F_\ell$ are obtained from a sum over an infinite number of OPE coefficients. However, by comparing the relative sizes of multiple partial wave coefficients, we will now show that we can also derive \emph{lower} bounds on the magnitudes of OPE coefficients of particular operators in the $\vep \times T$ OPE.

\begin{figure}[t!]
\centering
\includegraphics[width=.9\linewidth]{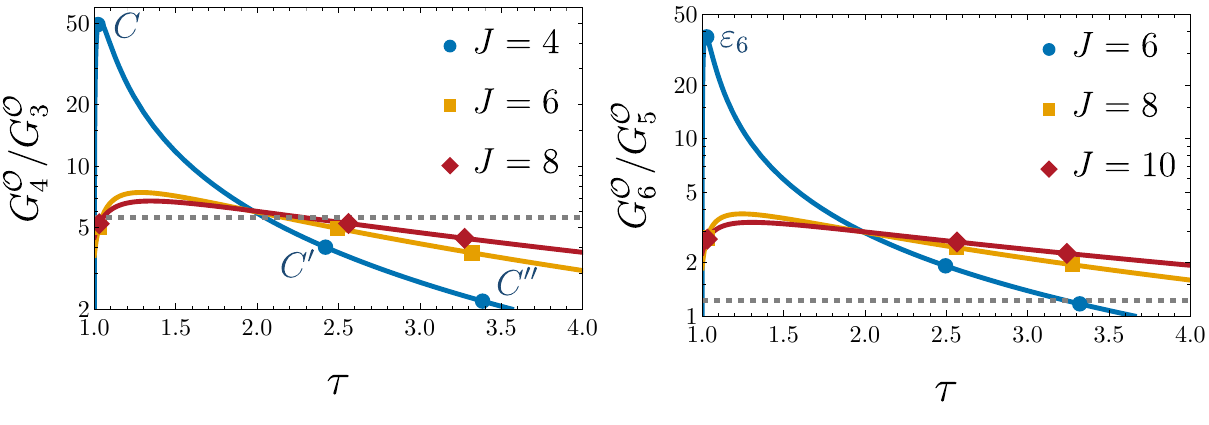}
\caption{\emph{Left:} Ratio of the $\ell=4$ conformal block coefficient $G^\Ocal_4$ to the $\ell=3$ conformal block coefficient $G^\Ocal_3$ as a function of the operator's twist $\tau \equiv \De-J$, for $J=4$ (blue), $J=6$ (yellow), and $J=8$ (red). \emph{Right:} Ratio of the $\ell=6$ conformal block coefficient $G^\Ocal_6$ to the $\ell=5$ conformal block coefficient $G^\Ocal_5$ as a function of the operator's twist $\tau$, for $J=6$ (blue), $J=8$ (yellow), and $J=10$ (red). In both plots, the solid lines show the general dependence on $\tau$ and each marker corresponds to a known operator in the $\vep \times T$ OPE for the 3d Ising CFT. The dashed gray lines indicate the thresholds from \eqref{eq:G3G4bound} and \eqref{eq:G5G6bound} that at least one operator must satisfy for the energy correlator in the background of $\vep$ to be positive.}
\label{fig:G3G4andG5G6}
\end{figure}

To start, let's again look at the bounds on $F_\ell$ in~\eqref{eq:FLBounds} obtained from the known OPE coefficient $\lambda_{\vep\hat{T}\hat{T}}$. The lower bound $F_4 \geq 0.744$ immediately tells us there must exist at least one operator with spin $J \geq 4$ in the $\vep \times T$ OPE, while the upper bound $F_3 \leq 0.132$ tells us that any such operator cannot have too large of a contribution to $F_3$. More precisely, \emph{there must exist at least one operator whose $\ell=4$ conformal block coefficient $G^\Ocal_4$ is sufficiently large compared to its $\ell=3$ coefficient $G^\Ocal_3$},
\be
\fr{G^\Ocal_4}{G^\Ocal_3} \geq \fr{0.744}{0.132} = 5.64.
\label{eq:G3G4bound}
\ee
The left plot in Figure~\ref{fig:G3G4andG5G6} shows this ratio of conformal block coefficients as a function of the twist $\tau \equiv \De - J$ for operators with spin $J=4,6,8$. Note that the conformal block coefficients depend on the scaling dimension of the external scalar operator, which in this plot we have set to the value of $\De_\vep$ from Table~\ref{tab:IsingData}. The markers in Figure~\ref{fig:G3G4andG5G6} indicate known operators in the 3d Ising CFT, and the gray line in the left plot shows the threshold from~\eqref{eq:G3G4bound} that at least one operator must satisfy.

As we can see from Figure~\ref{fig:G3G4andG5G6}, there is only one operator with $J=4$ whose ratio of conformal block coefficients is above this threshold: the minimal twist spin-$4$ operator $C$. Amazingly, the twist of $C$ is very close to the maximum for this ratio, though we are not aware of any reason why this needed to be the case. For our purposes here, all that matters is that $C$ is above the threshold while all other known spin-$4$ operators are below.

If $C$ is the \emph{unique} operator of \emph{any} spin whose conformal block coefficients are above this threshold, then we can place a lower bound on the magnitude of its OPE coefficient $|\lambda_{\vep\hat{T}C}|$. However, if we look more closely at Figure~\ref{fig:G3G4andG5G6} we see that as we increase $J$ the lowest and second-lowest twist operators of spin $J$ move close to the threshold, eventually crossing it at $J=14$. Fortunately, we can use the upper bound $F_5 \leq 0.360$ to significantly constrain the allowed OPE coefficients for these high-spin operators and still obtain the following lower bound (see appendix~\ref{app:LowerBounds} for details):
\be
\boxed{|\lambda_{\vep\hat{T}C}| \geq 0.218.}
\label{eq:LowerBoundC}
\ee
Note that this lower bound is rather conservative, as we can see by comparing with the known value in Table~\ref{tab:IsingData}. It would be very interesting to see if this bound could be improved by including expectations from the analytic bootstrap for the relative sizes of OPE coefficients for operators of different twist and spin~\cite{Fitzpatrick:2012yx,Komargodski:2012ek}.

In principle, we should be able to derive a similar lower bound on the OPE coefficient of the minimal twist operator $\vep_6$ at $J=6$. However, the partial wave coefficient bounds $F_5 \leq 0.360$ and $F_6 \geq 0.440$ from~\eqref{eq:FLBounds} are not sufficiently restrictive to compute a lower bound on a single operator. We can see this explicitly from the right plot in Figure~\ref{fig:G3G4andG5G6}, which compares the ratio of $G^\Ocal_6$ to $G^\Ocal_5$ for various operators in the 3d Ising CFT, compared to the threshold
\be
\fr{G^\Ocal_6}{G^\Ocal_5} \geq \fr{0.440}{0.360} = 1.22.
\label{eq:G5G6bound}
\ee
There are multiple operators above this threshold, which means we cannot currently obtain a lower bound on any individual OPE coefficient. However, it is interesting to note that $\vep_6$ is again located very close to the maximum for this ratio, such that there is a significant hierarchy between its ratio and that of all other operators. This suggests that if the bounds on $F_5$ and $F_6$ could be improved, we would be able to compute a lower bound on this operator's OPE coefficient.


\section{Bounds on OPE Coefficients in the 3d \texorpdfstring{$O(2)$}{O(2)} CFT}
\label{sec:O2Bounds}

While so far we have focused on the case of the 3d Ising CFT, the overall procedure outlined in section~\ref{sec:IsingBounds} yields bounds on OPE coefficients of the form $\lambda_{\phi T\Ocal}$ in any CFT in $d\geq3$, as long as the scaling dimensions of both the scalar operator $\phi$ and the spin-$J$ operator $\Ocal$ are known. These bounds can then be further strengthened by including the contribution of any known OPE coefficients to the partial wave coefficients $F_\ell$, as we did with $\lambda_{\vep TT}$ in the 3d Ising CFT.

To demonstrate our method in the context of a theory that is somewhat less well-understood than the Ising model, we now turn to the 3d $O(2)$ CFT. In particular, we will consider the bounds on OPE coefficients resulting from positivity of the energy correlator in the background of the lowest-dimension $O(2)$ singlet scalar operator: $s$.

\begin{table}[t!]
\begin{center}
\begin{tabular}{ |l|l|l|l||l|l| } 
\hline
& & & & & \\[-13pt]
$\Ocal$ & $\De$ & $J$ & $P$ & $|\lambda_{s\hat{T}\Ocal}|$ & Upper bound \\[1pt] 
\hline
$s$ & $1.51136(22)$~\cite{Chester:2019ifh} & $0$ & $+$ & $1.34710(20)$~\cite{Chester:2019ifh} & \\
\hline
& & & & & \\[-13pt]
$\hat{T}$ & $3$ & $2$ & $+$ & $0.762(16)$~\cite{Dey:2026cso} & 
\\
\hline
$s_4$ & $5.02548(41)$~\cite{Liu:2020tpf} & $4$ & $+$ &  & 
$0.399$ \\
\hline
$s_6$ & $7.03009(12)$~\cite{Liu:2020tpf} & $6$ & $+$ &  & 
$0.202$ \\
\hline
$s_8$ & $9.03311(18)$~\cite{Liu:2020tpf} & $8$ & $+$ &  & 
$0.102$ \\
\hline
\end{tabular}
\end{center}
\caption{Known primary operators $\Ocal$ with $\De < 10$ in the $s \times T$ OPE, with the associated scaling dimension $\De$, spin $J$, parity $P$, and OPE coefficient magnitude $|\lambda_{s \hat{T}\Ocal}|$ (if known), as well as the upper bound on $|\lambda_{s \hat{T}\Ocal}|$ computed using the method of section \ref{sec:IsingUpperBounds}. Note that $\hat{T} \equiv \fr{1}{\sqrt{c_T}} T$, as defined in~\eqref{eq:THatDef}.}
\label{tab:O2Data}
\end{table}

Table~\ref{tab:O2Data} lists the scaling dimension of $s$, as well as those of the known operators appearing in the $s \times T$ OPE, which have been computed with the conformal bootstrap~\cite{Chester:2019ifh,Liu:2020tpf}. From these scaling dimensions, we can again compute upper bounds on the associated OPE coefficients $\lambda_{s T\Ocal}$. Following our convention from~\eqref{eq:LambdaHatDef}, these are normalized by the central charge, which for the 3d $O(2)$ CFT is~\cite{Chester:2019ifh},
\be
c_T = 0.01793491(28).
\ee
As a concrete example, let's consider the OPE coefficient associated with the stress tensor $T$. Following the same approach as~\cite{Mecaj:2025ecl}, we immediately obtain the upper bound
\be
|\lambda_{s\hat{T}\hat{T}}| \leq \fr{1}{\sqrt{c_T \; G_2^T}} = 0.782.
\ee
Recently, this OPE coefficient was computed using fuzzy sphere regularization~\cite{Dey:2026cso}, yielding the value $\lambda_{s\hat{T}\hat{T}}=0.762(16)$, which is quite close to saturating the bound from positivity.

Just like for the 3d Ising CFT, we can now use this known OPE coefficient to place a lower bound on $F_2$. To be conservative, we will use the lowest value in the confidence interval for $\lambda_{s\hat{T}\hat{T}}$, resulting in the lower bound
\be
\boxed{F_2 \geq 0.912.}
\ee
Following the same procedure as section~\ref{sec:IsingPartialWave}, we can use this lower bound on $F_2$ to constrain the remaining $F_\ell$,
\be
\boxed{F_3 \leq 0.168, \quad F_4 \geq 0.677, \quad F_5 \leq 0.446, \quad F_6 \geq 0.368, \quad F_7 \leq 0.740.}
\ee
These bounds restrict the energy correlator in the background of $s$ to lie in a small region, corresponding to the red and yellow regions in Figure~\ref{fig:F2F3F4IsingO2}, compared to the full space of energy correlators. Though the bounds are somewhat weaker than in the Ising case (the red region), we still find that the allowed values for $F_\ell$ in the 3d $O(2)$ CFT must lie close to that of a free theory.

\begin{figure}[t!]
\centering
\includegraphics[width=0.55\linewidth]{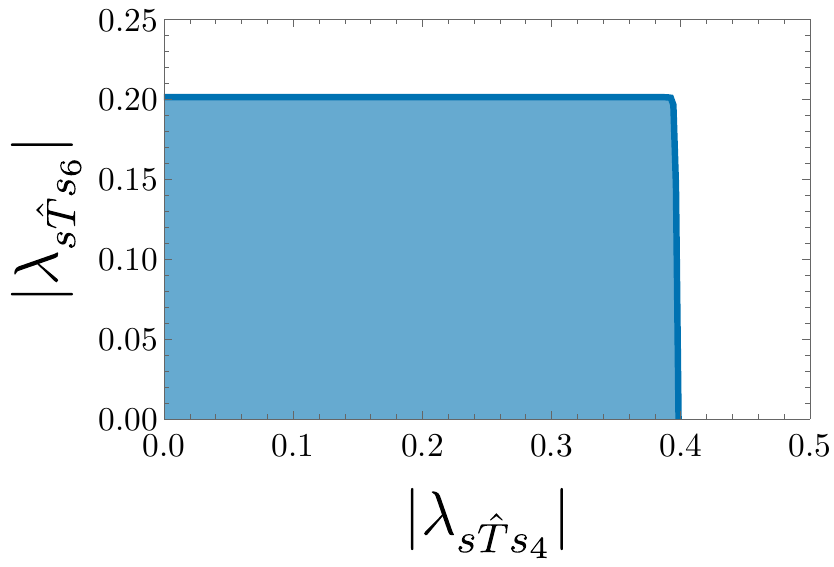}
\caption{Allowed values for the OPE coefficients of the lowest-twist operators with spins $J=4$ and $J=6$ in the 3d $O(2)$ CFT.}
\label{fig:O2Bounds}
\end{figure}

Given these improved bounds on the energy correlator partial wave coefficients, we can then implement the procedure from section~\ref{sec:IsingUpperBounds} to obtain stronger upper bounds on the OPE coefficients of the known operators in the $s \times T$ OPE. These are listed in the rightmost column of Table~\ref{tab:O2Data}. It would be very exciting to compare these bounds with the actual values for these OPE coefficients, once they are computed with methods like the conformal bootstrap or fuzzy sphere regularization.

We can also make joint plots showing the allowed region for multiple OPE coefficients, as shown in Figure~\ref{fig:O2Bounds} for the lowest-twist operators with spin $J=4$ and $J=6$. This is the $O(2)$ model analogue of the Ising plot in the upper left of Figure~\ref{fig:IsingBounds}. In both cases, we see that the bounds on the two OPE coefficients for operators are essentially independent, as the bound for $s_4$ is largely dominated by the constraint from $F_4$, while the bound for $s_6$ is dominated by the constraint from $F_6$.

Unfortunately, there is currently insufficient data on the spectrum of the 3d $O(2)$ CFT to place lower bounds on any OPE coefficients, as we did in section~\ref{sec:IsingLowerBounds} for the Ising CFT. Overall, it will be very interesting to see how the bounds from energy positivity improve as our knowledge of the spectrum of the $O(2)$ model and other CFTs develop in the future.


\section{EEC Positivity as a Bootstrap Problem}
\label{sec:Bootstrap}

So far, the bounds on OPE coefficients we have derived have been conservative in nature, assuming only a small number of known operators contribute to the lowest EEC partial wave coefficients $F_\ell$. We can generalize these initial bounds to a more systematic approach, which includes the contribution of an arbitrary number of operators to a large number of partial wave coefficients. This new approach, in the same spirit as the numerical conformal bootstrap~\cite{Poland:2018epd,Rychkov:2023wsd}, uses the ``crossing equation'' 
\be
\sum_{\Ocal} \lambda_{\phi T\Ocal}^2 \; G_{\ell}^\Ocal = \fr{1}{N_\ell} \int_0^1 dz \; \big(z(1-z)\big)^{\fr{d}{2}-2} F(z) \; \Chat_{\ell}(z).
\label{eq:Crossing}
\ee
These two decompositions of the partial wave coefficients $F_\ell$ are each a sum (or integral) of terms consisting of a theory-independent piece fixed by symmetry multiplied by a non-negative theory-dependent coefficient.

The fact that the conformal blocks $G_\ell^\Ocal$ are also non-negative allows us to truncate the left-hand side and still obtain rigorous upper bounds, such as those in sections~\ref{sec:IsingBounds} and \ref{sec:O2Bounds}, where we truncated to only the first few operators. These bounds can be generalized into the following semi-definite programming problem:
\be
\ba
\textrm{maximize} &&& \lambda_{\phi T \Ocal_1}^2 \\
\textrm{over the set} &&& \{\lambda_{\phi T \Ocal_i}^2 \; | \; i = 1, 2, \ldots, i_{\max}\} \\
\textrm{subject to} &&& \sum_i \lambda_{\phi T \Ocal_i}^2 \; G^{\Ocal_i}_\ell = F_\ell \quad (\ell =2,\ldots, \Lmax), \\
&&& H_{\Lmax}^0 \succeq 0, \\ 
&&& H_{\Lmax}^1 \succeq 0,\\ 
&&& H_{\Lmax-1}^0 - H_{\Lmax}^1 \succeq 0, \\ 
&&& H_{\Lmax-1}^1 - H_{\Lmax}^2 \succeq 0,
\ea
\label{eq:Primal}
\ee
where the Hankel matrices $H^\nmin_\nmax$ are defined in~\eqref{eq:HankelDef}.

\begin{table}[t!]
\begin{center}
\begin{tabular}{ |l|l|l| } 
\hline
& \\[-13pt]
$\Lmax$ & Upper bound $|\lambda_{\vep\hat{T}C}|$ & Upper bound $|\lambda_{\vep\hat{T}\vep_6}|$ \\[1pt] 
\hline
4 & 0.490 & \\
\hline
6 & 0.489 & 0.244 \\
\hline
8 & 0.483 & 0.243 \\
\hline
10 & 0.483 & 0.239 \\
\hline
20 & 0.482 & 0.238 \\
\hline
\end{tabular}
\end{center}
\caption{Upper bounds on the OPE coefficients $|\lambda_{\vep\hat{T} C}|$ and $|\lambda_{\vep\hat{T}\vep_6}|$ from the primal approach outlined in~\eqref{eq:Primal} for different values of $\Lmax$. The inputs for computing these bounds were the known scaling dimensions of the leading twist families $[\sigma\sigma]_0$, $[\sigma\sigma]_1$, and $[\vep\vep]_0$~\cite{Simmons-Duffin:2016wlq} and the known value for the OPE coefficient $\lambda_{\vep TT}$~\cite{Chang:2024whx}.}
\label{tab:PrimalBounds}
\end{table}

For example, Table~\ref{tab:PrimalBounds} shows the resulting upper bounds on $|\lambda_{\vep\hat{T}C}|$ and $|\lambda_{\vep\hat{T}\vep_6}|$ obtained by including all known low-twist operators with spin $J \leq 40$ in the $\vep \times T$ OPE, whose scaling dimensions were computed in~\cite{Simmons-Duffin:2016wlq}, as well as the known value for $\lambda_{\vep TT}$ from~\cite{Chang:2024whx}. As we can see, including this much larger set of known scaling dimensions provides only percent-level improvement to our original bounds in Table~\ref{tab:IsingData}. To significantly improve these bounds, some additional input must be added to this analysis, such as more known OPE coefficients or further constraints beyond positivity and unitarity.

The method outlined in~\eqref{eq:Primal} is the so-called ``primal'' approach to such bounds. A more systematic method of accounting for an arbitrary number of operators is the ``dual'' approach (see, e.g.,~\cite{Afkhami-Jeddi:2019zci} for a more detailed presentation), which we can formulate by first rewriting the crossing equation~\eqref{eq:Crossing} into a vanishing sum of vectors,
\be
\mathbf{G}_{\phi} + \sum_\Ocal \lambda_{\phi T\Ocal}^2 \; \mathbf{G}_\Ocal + \fr{1}{N_0} \int_0^1 dz \; \big(z(1-z)\big)^{\fr{d}{2}-2} F(z) \; \mathbf{C}(z) = 0,
\label{eq:CrossingVec}
\ee
where $\mathbf{G}_{\phi},\mathbf{G}_\Ocal,\mathbf{C}(z) \in \mathbb{R}^{\ell_{\rm max}+1}$ are defined as
\be
\mathbf{G}_{\phi} \equiv \left( \begin{array}{c} 1 \\ 0 \\ 0 \\ \vdots \\ 0 \end{array} \right), \quad \mathbf{G}_\Ocal \equiv \left( \begin{array}{c} 0 \\ 0 \\ G_2^\Ocal \\ \vdots \\ G_\Lmax^\Ocal \end{array} \right), \quad \mathbf{C}{(z)} \equiv -\left( \begin{array}{c} 1 \\ 1-2z \\ \fr{N_0}{N_2} \; \Chat_2(1-2z) \\ \vdots \\ \fr{N_0}{N_\Lmax} \; \Chat_\Lmax(1-2z) \end{array} \right).
\ee
We can then make some set of assumptions about the operators $\Ocal$ appearing in the OPE $\phi \times T$. The vectors $\mathbf{G}_\Ocal$ and $\mathbf{C}$ allowed by these assumptions, when combined with non-negative coefficients, span a convex cone. If there exists a hyperplane, which we can specify by its normal vector $\boldsymbol{\alpha}$, such that the convex cone is entirely on one side of that plane, then the crossing equation~\eqref{eq:CrossingVec} is inconsistent, which means our set of assumptions is excluded. Semi-definite programming solvers such as {\tt sdpb}~\cite{Simmons-Duffin:2015qma,Landry:2019qug} either find such $\boldsymbol{\alpha}$ or prove that no such $\boldsymbol{\alpha}$ exists.

As a concrete example of this procedure, the setup for placing upper/lower bounds on the OPE coefficient of the minimal-twist operator $\Ocal_{\min}$ of a particular spin is the following:
\be
\ba
\textrm{maximize} &&& \boldsymbol{\alpha} \cdot \mathbf{G}_{\Ocal_{\min}} \\
\textrm{subject to} &&& \boldsymbol{\alpha} \cdot \mathbf{G}_\phi = \mp 1, \\
&&& \boldsymbol{\alpha} \cdot \mathbf{G}_\Ocal \geq 0 && \forall \,\, J \geq 2, \, \De \geq \De_\gap(J), \\ 
&&& \boldsymbol{\alpha} \cdot \mathbf{C}(z) \geq 0 && \forall \,\, 0 \leq z \leq 1,
\ea
\label{eq:Dual}
\ee
where $\De_\gap(J)$ is our assumption for the lowest dimension of any operator (excluding $\Ocal_{\min}$) at each spin $J$. To see how this leads to a bound on the OPE coefficient, note that we can relax~\eqref{eq:CrossingVec} to the following inequality
\be
\ba
\boldsymbol{\alpha} \cdot \mathbf{G}_{\phi} + \lambda_{\phi T\Ocal_{\min}}^2 \! \boldsymbol{\alpha} \cdot \mathbf{G}_{\Ocal_{\min}} &= -\sum_{\Ocal\neq \Ocal_{\min}} \lambda_{\phi T\Ocal}^2 \; \boldsymbol{\alpha} \cdot \mathbf{G}_\Ocal - \fr{1}{N_0} \int_0^1 dz \; \big(z(1-z)\big)^{\fr{d}{2}-2} F(z) \; \boldsymbol{\alpha} \cdot \mathbf{C}(z) \\
&\leq 0.
\ea
\ee
We have isolated the contributions from $\phi$ and $\Ocal_{\min}$, whose sum must be non-positive to be consistent with~\eqref{eq:CrossingVec}, if an $\boldsymbol{\alpha}$ satisfying the conditions in~\eqref{eq:Dual} exists. If $\boldsymbol{\alpha} \cdot \mathbf{G}_\phi = - 1$, we automatically obtain the upper bound,
\be
\lambda_{\phi T\Ocal_{\min}}^2 \leq \fr{1}{\boldsymbol{\alpha} \cdot \mathbf{G}_{\Ocal_{\min}}},
\ee
which is clearly optimized by maximizing $\boldsymbol{\alpha} \cdot \mathbf{G}_{\Ocal_{\min}}$. If instead we choose $\boldsymbol{\alpha} \cdot \mathbf{G}_\phi = 1$, then $\boldsymbol{\alpha} \cdot \mathbf{G}_{\Ocal_{\min}}$ must be negative to satisfy~\eqref{eq:CrossingVec}, resulting in the lower bound
\be
\lambda_{\phi T\Ocal_{\min}}^2 \geq \fr{1}{|\boldsymbol{\alpha} \cdot \mathbf{G}_{\Ocal_{\min}}|},
\ee
which is again optimized by maximizing $\boldsymbol{\alpha} \cdot \mathbf{G}_{\Ocal_{\min}}$ (i.e.~minimizing $|\boldsymbol{\alpha} \cdot \mathbf{G}_{\Ocal_{\min}}|$).

It would be interesting to explore this ``EEC bootstrap'' approach further, to determine the extent to which positivity constrains the spectrum and OPE coefficients of individual CFTs, such as the 3d Ising and $O(2)$ models. Based on the preliminary results in Table~\ref{tab:PrimalBounds}, our naive expectation is that while this approach can derive non-trivial bounds on OPE coefficients similar to those in sections~\ref{sec:IsingBounds} and \ref{sec:O2Bounds}, positivity alone is insufficient to place high-precision bounds on CFT data. However, if we were to instead consider energy correlators in the background of spinning external states, then we could combine EEC positivity with the superconvergence sum rules resulting from the vanishing commutator of energy flow operators~\cite{Kologlu:2019bco,Caron-Huot:2020adz,Chang:2023szz},\footnote{The reason this sum rule requires an external state $|\psi\>$ with spin to result in nontrivial constraints is that the conformal blocks in the scalar case are manifestly symmetric under the exchange $\hat{n}_1 \lra \hat{n}_2$, so~\eqref{eq:Superconvergence} is automatically satisfied.}
\be
\<\psi|\comm{\Ecal(\hat{n}_1)}{\Ecal(\hat{n}_2)}|\psi\> = 0.
\label{eq:Superconvergence}
\ee
The conformal block decomposition of this commutator would result in a genuine crossing equation with CFT data on both sides, which we expect would place much stronger constraints on both the operator spectrum and OPE coefficients.


\section{Discussion}
\label{sec:Discussion}

In this work, we have explored the universal constraints placed by energy positivity on the partial wave coefficients of energy correlators in general unitary QFTs in $d\geq3$. We have shown that the partial wave coefficients for two-point energy correlators are forced by positivity and unitarity to lie within a bounded region: the EEC-hedron. For the particular case of CFTs, these positivity constraints lead to bounds on the allowed values for OPE coefficients involving the stress-energy tensor. We have demonstrated the use of these positivity bounds in two examples, the 3d Ising and $O(2)$ CFTs, deriving new constraints on the allowed values of multiple unknown OPE coefficients. Using data from the conformal bootstrap and fuzzy sphere regularization, we have also shown that energy correlators in the background of the lowest-dimension global singlet scalar operators for both theories are close in structure to those of a free theory.

There are many interesting directions for future research expanding on our results:

{\bf{Bounding OPE coefficients in other CFTs:}} The method we have presented here can be applied to any other CFT, provided there is at least some knowledge of the spectrum of scaling dimensions. One obvious candidate for study is $\Ncal=4$ supersymmetric Yang-Mills, where there is a wealth of data from perturbation theory~\cite{Kotikov:2004er,Eden:2012rr,Chicherin:2015edu}, the conformal bootstrap~\cite{Beem:2013qxa,Beem:2016wfs,Henriksson:2017eej,Chester:2021aun,Caron-Huot:2022sdy,Chester:2023ehi,Caron-Huot:2024tzr}, integrability~\cite{Gromov:2013pga,Gromov:2014bva,Basso:2015zoa,Basso:2015eqa,Bargheer:2017nne,Bargheer:2018jvq,Gromov:2023hzc}, and localization~\cite{Binder:2019jwn,Chester:2019jas,Chester:2020dja,Chester:2020vyz,Alday:2023pet}. A significant step in this direction was taken in~\cite{Dempsey:2025yiv}, which used this $\Ncal=4$ SYM data to place highly restrictive bounds on the energy correlator in the background of half-BPS operators. It would be very interesting to use these constraints on the energy correlator to derive bounds on a large number of OPE coefficients at finite $N$ and finite coupling.

{\bf{Optimal bounds for finite number of partial waves:}} The interplay between energy positivity and unitarity in constraining the allowed space of partial wave coefficients $F_\ell$ is rather nontrivial, such that in principle one must consider an infinite number of constraints to determine the optimal bounds for a finite set of coefficients (see appendix~\ref{app:Lmax} for details). Phrased in geometric language, this problem amounts to determining the intersection between the convex hull of moment curves (i.e.~the space of $F_\ell$ computed from a positive measure) and the individual hyperplanes where each $F_\ell=0$. While in practice one can obtain an outer approximation of the allowed region from a finite set of constraints, it would be useful to have a method for directly determining the precise region of intersection without the need to consider infinitely large Hankel matrices.

{\bf{Additional constraints for physical theories:}} In defining the space of allowed energy correlators in general unitary QFTs, we have only made two assumptions: both the function $F(z)$ and its partial wave coefficients $F_\ell$ must be non-negative. However, it is unclear what (if any) additional constraints must be imposed on $F(z)$ to ensure it arises from a local QFT. In other words, are the bounds we have computed on $F_\ell$ saturated by energy correlators in physical theories, or are there additional universal constraints that will further reduce the allowed region? If the positivity bounds are saturated by actual QFTs, the most plausible candidates are free theories, as the extremal functions $F(z)$ along the boundary correspond to a sum of delta functions.

For the specific case of CFTs, we have imposed one additional constraint: the partial wave coefficient $F_1$ must vanish, as required by momentum conservation. It seems very likely, however, that additional constraints need to be imposed to ensure consistency with fundamental properties of CFTs, such as conformal invariance and convergence of the OPE. It would be exciting to determine these additional constraints and potentially further improve the bounds on OPE coefficients computed in this work.

{\bf{Higher-point energy correlators:}} Here we have focused on two-point energy correlators, which are only a function of a single variable $z = \half(1-\cos\theta)$. Higher-point energy correlators are instead non-negative functions of multiple variables $z_i$, with a more complicated partial wave expansion. It would be useful to determine the analogous constraints placed by positivity and unitarity on the partial wave coefficients of higher-point energy correlators. This would be particularly interesting for the case of CFTs, where one can also generalize the conformal block calculation of~\cite{Mecaj:2025ecl} to higher-point correlators, in order to constrain general OPE coefficients involving the stress-energy tensor.

{\bf{Positivity for generalized detectors:}} While in this work we have focused on the energy flow operator $\Ecal(\hat{n})$, it is not the only known positive detector operator. In fact, the proof of the ANEC in~\cite{Hartman:2016lgu} was shown to generalize to an infinite set of detectors built from the lowest-twist operator at every even spin. It would be very interesting to extend the EEC-hedron analysis presented here to correlators built from these more general detectors. This would be particularly powerful in CFTs, where one could derive bounds on a much larger class of OPE coefficients, not just those involving the stress-energy tensor.

{\bf{Energy correlators with slightly broken higher-spin symmetry:}} One of the more intriguing results of our analysis is the proximity of the values for $F_\ell$ in the 3d Ising and $O(2)$ CFTs to that of a free theory. This is perhaps not too surprising, as the $F_\ell$ appear to receive their dominant contribution from operators in the leading Regge trajectory, which in these particular theories are well-described by slightly broken higher-spin currents~\cite{Maldacena:2012sf,Alday:2015ota}. It would therefore be interesting to compute the general behavior of energy correlators in the presence of slightly broken higher-spin symmetry, as a simple model for the Ising and $O(N)$ CFTs.

{\bf{Positivity bounds on QFT parameters:}} In this work we were able to convert positivity constraints on energy correlators into rigorous bounds on the allowed values for OPE coefficients in individual CFTs. This approach hinged on obtaining a representation of the energy correlator, i.e.~the conformal block decomposition, which was \emph{not} manifestly positive. It would be exciting if an analogous approach could be applied to more general QFTs, in order to derive bounds on couplings or the coefficients of higher-dimensional operators in individual theories. One potential first step in this direction would be to connect the positivity bounds considered in this work with the ANEC-based derivations of the $c$- and $a$-theorems in~\cite{Hartman:2023qdn,Hartman:2023ccw,Hartman:2024xkw}.

{\bf{Connection to numerical bootstrap:}} Currently the most powerful method of constraining CFT data is the numerical conformal bootstrap, which uses crossing symmetry of correlation functions of local operators to derive rigorous bounds on scaling dimensions and OPE coefficients (see~\cite{Poland:2018epd,Rychkov:2023wsd} for reviews). This approach is applied in Euclidean signature, with all operators inserted at spacelike separation. What is intriguing about the CFT constraints derived in this work is that they do not rely on crossing symmetry and probe a different kinematic regime, with light-ray operator detectors inserted at timelike separation from a local operator source. For OPE coefficients of low-dimension operators, we have seen that results from the conformal bootstrap are all consistent with our positivity bounds, but it would be interesting to see whether this continues to hold for operators with larger scaling dimension and spin. One particularly exciting prospect would be to combine the two methods, using EEC-hedron bounds as a built-in constraint on the extremal solutions discovered with the conformal bootstrap.

{\bf{Bootstrapping CFTs with spinning energy correlators:}} In section~\ref{sec:Bootstrap} we showed how EEC positivity can be rephrased as a bootstrap problem, where assumptions about the operator spectrum and OPE coefficients can be tested for consistency with positivity and unitarity. Based on our initial work in this direction, it appears that applying these constraints to energy correlators in the background of scalar operators is insufficient to isolate individual physical CFTs. However, it would be very interesting to generalize this approach to external states with spin, where two new features arise. First, the energy correlator becomes a positive semi-definite matrix with entries labeled by the various external polarizations~\cite{Riembau:2025isw}. Second, the superconvergence sum rules of~\cite{Kologlu:2019bco,Caron-Huot:2020adz,Chang:2023szz} introduce a new crossing equation relating two distinct conformal block decompositions of the energy correlator. We expect these new constraints to lead to significantly more restrictive bounds than in the scalar case, particularly when the external state is created by the stress-energy tensor itself, potentially allowing us to determine the spectrum and OPE coefficients of individual CFTs directly from fundamental properties of energy correlators.


\section*{Acknowledgments}    

It is a pleasure to thank Cyuan-Han Chang, Clay Cordova, Rajeev Erramilli, Kara Farnsworth, Denis Karateev, Wei Li, Brett Oertel, Miguel Paulos, John Stout, and Francesco Riva for invaluable discussions. BM thanks the Erwin-Schrödinger International Institute for Mathematics and Physics at the University of Vienna for partial support during the Programme ``New Paradigms for Harnessing Quantum Field Theory at Colliders", July 27 -- August 28, 2026. BM is supported by the US Department of Energy through the Los Alamos National Laboratory and by LANL’s Laboratory Directed Research and Development (LDRD) program with project numbers 20230872PRD4 and 20250214ER. Los Alamos National Laboratory is operated by Triad National Security, LLC, for the National Nuclear Security Administration of the U.S. Department of Energy (Contract No.~89233218CNA000001). IM is supported by the DOE Early Career Award DE-SC0025581, and the Sloan Foundation. MW is supported by the Royal Society University Research Fellowship URF{\textbackslash}R1{\textbackslash}221905. This research was also supported in part by grant NSF PHY-2309135 to the Kavli Institute for Theoretical Physics (KITP).


\appendix

\section{Dependence of Partial Wave Bounds on \texorpdfstring{$\Lmax$}{Lmax}}
\label{app:Lmax}

\begin{figure}[t!]
\centering
\includegraphics[width=.55\linewidth]{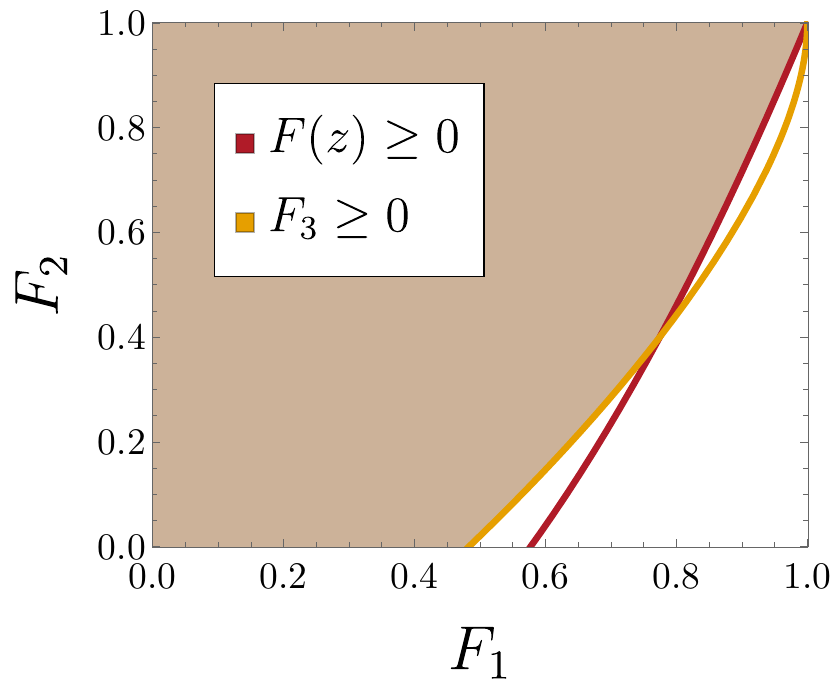}
\caption{Lower bound on $F_2$ as a function of $F_1$ for 4d QFTs obtained solely from positivity of $F(z)$ (red) and from additionally requiring positivity of $F_3$ (yellow). The brown region satisfying both lower bounds corresponds to the $\Lmax=3$ bounds in the left plot of Figure~\ref{fig:F1F2}.}
\label{fig:F3Bound}
\end{figure}

In this appendix we discuss how the allowed values for a given partial wave coefficient $F_\ell$ depend on the constraints for higher coefficients $F_{\ell'}$ with $\ell' > \ell$. For concreteness, let's consider the allowed values for $F_1$ and $F_2$ for a general QFT in $d=4$. Following~\eqref{eq:PositivityBounds}, we know these coefficients must satisfy the positivity constraints
\be
\begin{pmatrix} 1 & \half(1-F_1) \\ \half(1-F_1) & \fr{1}{6}(2-3F_1+F_2) \end{pmatrix} \succeq 0, \quad \half(1-F_1) \geq 0, \quad \half(1+F_1) \geq 0, \quad \fr{1}{6}(1-F_2) \geq 0,
\ee
as well as the unitarity constraints from~\eqref{eq:UnitarityBounds}
\be
F_1 \geq 0, \quad F_2 \geq 0.
\ee
Note that we've already imposed the energy conservation requirement $F_0 = 1$ from~\eqref{eq:EnergyCons}. Combining these together, we find
\be
0 \leq F_1 \leq 1, \quad 0 \leq F_2 \leq 1, \quad F_2 \geq \half (3F_1^2 - 1),
\ee
the last of which is shown by the red line in Figure~\ref{fig:F3Bound}. The region bounded by these three constraints corresponds to the $\Lmax=2$ bounds in the left plot of Figure~\ref{fig:F1F2}.

If we now include $F_3$ in this analysis, we obtain the additional constraints
\be
\ba
&\hspace{0.5cm} \begin{pmatrix} \half(1-F_1) & \fr{1}{6}(2-3F_1+F_2) \\ \fr{1}{6}(2-3F_1+F_2) & \fr{1}{20}(5-9F_1+5F_2-F_3) \end{pmatrix} \succeq 0, \\
&\begin{pmatrix} \half(1+F_1) & \fr{1}{6}(1-F_2) \\ \fr{1}{6}(1-F_2) & \fr{1}{60}(5-3F_1-5F_2+3F_3) \end{pmatrix} \succeq 0, \quad F_3 \geq 0.
\ea
\ee
From these we can derive a new constraint on the allowed values for $F_1$ and $F_2$,
\be
\fr{5-6F_1-9F_1^2+5F_2+15F_1 \; F_2-10F_2^2}{9(1-F_1)} \geq F_3 \geq 0,
\ee
or equivalently,
\be
F_2 \geq \fr{1}{4}(3F_1+1)-\fr{3}{4}\sqrt{\fr{1}{5}(1-F_1)(3F_1+5)},
\ee
which corresponds to the yellow line in Figure~\ref{fig:F3Bound}. As we can see, for $\fr{1}{3}(\sqrt{6}-1) \leq F_1 \leq \sqrt{3/5}$ this new constraint is actually stronger than the one derived solely from positivity of $F(z)$. In other words, positivity of $F_3$ (i.e.~unitarity) further reduces the allowed region for $F_1$ and $F_2$, leading to the brown region in Figure~\ref{fig:F3Bound}, which is the same as the $\Lmax=3$ bounds in the left plot of Figure~\ref{fig:F1F2}.

\begin{figure}[t!]
\centering
\includegraphics[width=.9\linewidth]{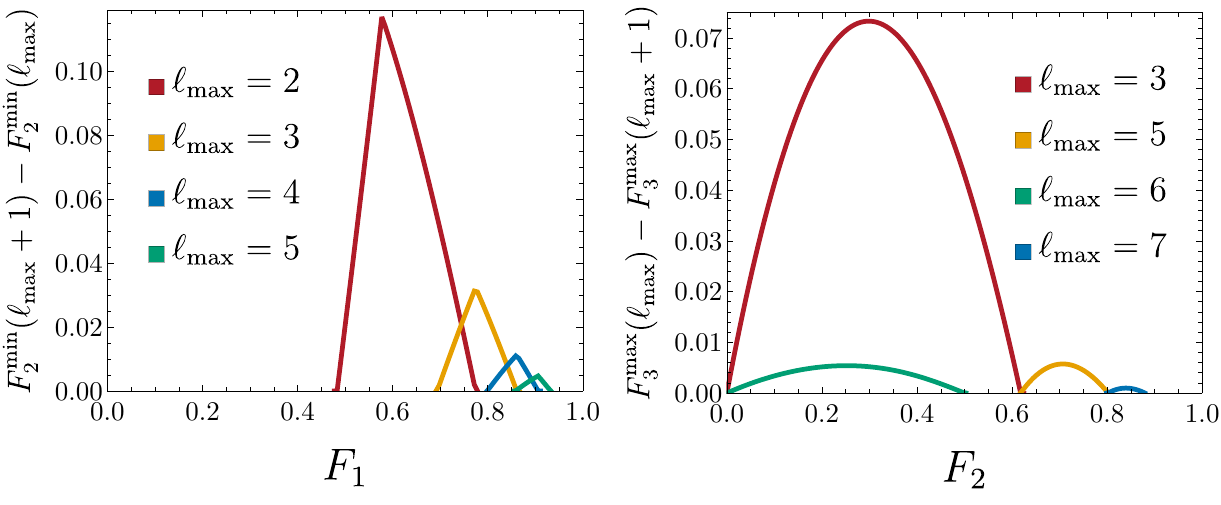}
\caption{\emph{Left:} Increase in the lower bound on $F_2$ in 4d QFTs when moving from $\Lmax$ to $\Lmax+1$ as a function of $F_1$ for $\Lmax = 2$ (red), $3$ (yellow), $4$ (blue), and $5$ (green). \emph{Right:} Decrease in the upper bound on $F_3$ in 3d CFTs when moving from $\Lmax$ to $\Lmax+1$ as a function of $F_2$ for $\Lmax = 3$ (red), $5$ (yellow), $6$ (green), and $7$ (blue).}
\label{fig:F2F3Shifts}
\end{figure}

To better understand why positivity of $F_3$ modifies the bounds, let's look more carefully at the function $F(z)$ that saturates the original positivity bounds at $\Lmax=2$. One can show that the unique positive function with $F_0 = 1$ and $F_2 = \half (3F_1^2 - 1)$ is a single delta function,
\be
F(z) = \delta\Big(z-\tfr{1}{2}(1-F_1)\Big).
\ee
For $F_1 \geq \sqrt{1/3}$, this function gives the lowest possible value for $F_2$ allowed by the $F(z) \geq 0$ bounds in Figure~\ref{fig:F3Bound}. However, if we now evaluate $F_3$ for this extremal function, we find
\be
F_3 \equiv \int_0^1 dz \; F(z) \; (1-12z+30z^2-20z^3) = \half F_1(5F_1^2-3),
\ee
which is \emph{negative} for $F_1 < \sqrt{3/5}$. In other words, the extremal solution at $\Lmax=2$ violates unitarity of $F_3$ for a finite range of $F_1$, hence the lower bound on $F_2$ for those values of $F_1$ must increase at $\Lmax=3$.

If we repeat this analysis for $F_4$ and higher, we find that each $F_\ell$ computed from the extremal function is negative for some finite range of $F_1$ values. To satisfy unitarity, the lower bound on $F_2$ must increase each time we increase $\Lmax$, as shown in the left plot of Figure~\ref{fig:F2F3Shifts}. As we can see, this correction to the bounds occurs over a finite range of $F_1$ values and quickly becomes negligible as we move to large values of $\Lmax$.

This same behavior occurs for bounds on CFTs, where we impose the additional constraint $F_1 = 0$ from~\eqref{eq:MomCons}. For example, the right plot in Figure~\ref{fig:F2F3Shifts} shows the decrease in the upper bound on $F_3$ at each $\Lmax$ in 3d CFTs. Again, the correction to the bound is limited to a finite range of $F_2$ and quickly becomes negligible at large $\Lmax$.


\section{Parity-Odd Conformal Blocks}
\label{app:OddBlocks}

As mentioned in section~\ref{sec:Recap}, in $d=3$ there is an allowed parity-odd tensor structure for the three-point functions $\<\phi T \Ocal\>$, in addition to the usual parity-even one. In theories with parity symmetry, such as the Ising and $O(2)$ CFTs, only one of these two structures will contribute to a given three-point function. If we assume that $\phi$ is parity-even, like $\vep$ in the Ising CFT and $s$ in the $O(2)$ CFT, then the correlator $\<\phi T \Ocal^+\>$ with a parity-even operator $\Ocal^+$ is given by the parity-even tensor structure~\eqref{eq:OPEDefEven} and the correlator with a parity-odd operator $\Ocal^-$ is given by the parity-odd tensor structure~\eqref{eq:OPEDefOdd}.

The resulting energy correlator conformal blocks for both of these cases were computed in~\cite{Mecaj:2025ecl}, but in this appendix we show that the conformal blocks $G_{\Ocal^-}(z)$ for parity-odd operators are actually proportional to the blocks $G_{\Ocal^+}(z)$ for parity-even operators. This relation between parity-even and parity-odd blocks was not manifest in the derivation presented in~\cite{Mecaj:2025ecl}. However, this result is perhaps unsurprising, as the conformal block $G_{\Ocal^\pm}(z)$ in both cases must be a parity-even eigenfunction of the quadratic Casimir of the conformal group,\footnote{While the individual three-point functions $\<\phi T \Ocal^\pm\>$ are even/odd under parity, the conformal blocks $G_{\Ocal^\pm}(z)$ are proportional to $|\<\phi T \Ocal^\pm\>|^2$ and are therefore parity-even in both cases.} which presumably can be shown to be unique once the scaling dimension $\De$ and spin $J$ of $\Ocal$ are specified. It would be interesting to find a derivation of the blocks which makes this structure manifest.

In general, the conformal blocks for both parity-even and parity-odd operators can be expanded in terms of partial waves,
\be
G_{\Ocal^\pm}(z) = \sum_{\ell=2}^J G_\ell^{\Ocal^\pm} \; \Chat_\ell(1-2z).
\ee
In~\cite{Mecaj:2025ecl}, the partial wave coefficients $G_\ell^{\Ocal^\pm}$ were computed from three-point function coefficients $C_\ell^{\Ocal^\pm}$ via the relation\footnote{The coefficients $C^{\Ocal^\pm}_\ell$ here correspond to $\Ccal_{\phi\Ecal\Ocal}^{(\ell)}$ in (3.25) and $\Ccal_{\phi\Ecal\Ocal}^{(\ell^-)}$ in (3.35) of~\cite{Mecaj:2025ecl}.}
\be
G_\ell^{\Ocal^\pm} \equiv \fr{2^{2\De_\phi+2\De-J-8} (2J)! \; \G(\De_\phi-\half) \; \G(\De_\phi) \; \G(\De-\half) \; \G(\De+J) \; \G(\De-J-1)}{\pi^5 \; (J-\ell)! \; (J+\ell)! \; (\De-1)_\ell \; \G(\De-\ell-1)} \; \big|C_\ell^{\Ocal^\pm}\big|^2.
\label{eq:CtoG}
\ee
To show that the parity-odd blocks are proportional to the parity-even blocks, it is therefore sufficient to show that $C^{\Ocal^-}_\ell$ is proportional to $C^{\Ocal^+}_\ell$ with an $\ell$-independent coefficient.

We'll start with the case $J=2$ as an introductory example, though this case is trivial as there is only one coefficient $C^{\Ocal^\pm}_2$. Using the results of~\cite{Mecaj:2025ecl}, we obtain
\be
\ba
C_2^{\Ocal^+} &= \fr{24\pi^4}{\G\big(\fr{\De_\phi-\De+7}{2}\big) \; \G\big(\fr{\De-\De_\phi+3}{2}\big) \; \G(\De+\De_\phi)}, \\
C_2^{\Ocal^-} &= \fr{12\pi^4}{\G\big(\fr{\De_\phi-\De+6}{2}\big) \; \G\big(\fr{\De-\De_\phi+4}{2}\big) \; \G(\De+\De_\phi)},
\ea
\ee
with the resulting ratio
\be
\fr{C_2^{\Ocal^-}}{C_2^{\Ocal^+}} = \fr{\G\big(\fr{\De_\phi-\De+7}{2}\big) \; \G\big(\fr{\De-\De_\phi+3}{2}\big)}{2\G\big(\fr{\De_\phi-\De+6}{2}\big) \; \G\big(\fr{\De-\De_\phi+4}{2}\big)}.
\ee

Next, we can consider the case $J=3$, with the resulting coefficients
\be
\ba
C_2^{\Ocal^+} &= \fr{4\pi^4\big(\De(\De-3)-\De_\phi(\De_\phi-3)-6\big)}{\G\big(\fr{\De_\phi-\De+8}{2}\big) \; \G\big(\fr{\De-\De_\phi+4}{2}\big) \; \G(\De+\De_\phi+1)}, \\
C_2^{\Ocal^-} &= \fr{2\pi^4\big(\De(\De-3)-\De_\phi(\De_\phi-3)-6\big)}{\G\big(\fr{\De_\phi-\De+7}{2}\big) \; \G\big(\fr{\De-\De_\phi+5}{2}\big) \; \G(\De+\De_\phi+1)}, \\
C_3^{\Ocal^+} &= \fr{96\pi^4(\De+1)}{\G\big(\fr{\De_\phi-\De+8}{2}\big) \; \G\big(\fr{\De-\De_\phi+4}{2}\big) \; \G(\De+\De_\phi+1)}, \\
C_3^{\Ocal^-} &= \fr{48\pi^4(\De+1)}{\G\big(\fr{\De_\phi-\De+7}{2}\big) \; \G\big(\fr{\De-\De_\phi+5}{2}\big) \; \G(\De+\De_\phi+1)},
\ea
\ee
whose ratios are independent of $\ell$,
\be
\fr{C_2^{\Ocal^-}}{C_2^{\Ocal^+}} = \fr{C_3^{\Ocal^-}}{C_3^{\Ocal^+}} = \fr{\G\big(\fr{\De_\phi-\De+8}{2}\big) \; \G\big(\fr{\De-\De_\phi+4}{2}\big)}{2\G\big(\fr{\De_\phi-\De+7}{2}\big) \; \G\big(\fr{\De-\De_\phi+5}{2}\big)}.
\ee

One can proceed to higher values of $J$, each time finding that the ratios of parity-odd to parity-even coefficients are independent of $\ell$, with the general result
\be
\fr{C_\ell^{\Ocal^-}}{C_\ell^{\Ocal^+}} = \fr{\G\big(\fr{\De_\phi-\De+J+5}{2}\big) \; \G\big(\fr{\De+J-\De_\phi+1}{2}\big)}{2\G\big(\fr{\De_\phi-\De+J+4}{2}\big) \; \G\big(\fr{\De+J-\De_\phi+2}{2}\big)}.
\ee
Using~\eqref{eq:CtoG}, we can then obtain the relation between parity-odd and parity-even conformal block coefficients presented in~\eqref{eq:OddEvenRelation}. One important consequence of the overall proportionality constant is a shift in the zeroes of these conformal blocks, from $\tau \equiv \De-J=\De_\phi+5+2n$ for parity-even blocks to $\tau=\De_\phi+4+2n$ for parity-odd blocks.


\section{Calculation of OPE Coefficient Lower Bounds}
\label{app:LowerBounds}

In this appendix we provide more details of the procedure for obtaining the lower bounds on OPE coefficients presented in section~\ref{sec:IsingLowerBounds}. In general, this calculation can be rephrased as the following linear optimization problem:
\be
\ba
\textrm{minimize} &&& \lambda_{\vep T \Ocal_1}^2 \\
\textrm{subject to} &&& \sum_i \lambda_{\vep T \Ocal_i}^2 \; G^{\Ocal_i}_{\ell_1} \geq F_{\ell_1}^{\min}, \\
&&& \sum_i \lambda_{\vep T \Ocal_i}^2 \; G^{\Ocal_i}_{\ell_2} \leq F_{\ell_2}^{\max}, \\
&&& \ldots
\ea
\ee
where the various constraints come from the upper and lower bounds on the partial wave coefficients $F_\ell$ computed in section~\ref{sec:IsingPartialWave}. There are many standard algorithms for solving problems of this form (see, e.g.~\cite{Bertsimas:1997}), so here we will mostly focus on how non-trivial lower bounds arise rather than efficient methods for numerically computing them.

As a concrete example, let's consider the case with only two constraints: an upper bound on $F_3$ and a lower bound on $F_4$. These two constraints lead to a bound on the ratio of $F_4$ to $F_3$,
\be
\fr{F_4}{F_3} = \fr{\sum_i \lambda_{\vep T \Ocal_i}^2 \; G^{\Ocal_i}_4}{\sum_i \lambda_{\vep T \Ocal_i}^2 \; G^{\Ocal_i}_3} \geq \fr{F_4^{\min}}{F_3^{\max}},
\label{eq:F4F3Bound}
\ee
which we can rewrite as
\be
\fr{F_4}{F_3} = \sum_i \bigg(\fr{\lambda_{\vep T \Ocal_i}^2 \; G^{\Ocal_i}_3}{F_3} \cdot \fr{G^{\Ocal_i}_4}{G^{\Ocal_i}_3} \bigg) \geq \fr{F_4^{\min}}{F_3^{\max}}.
\ee
In other words, the ratio of $F_4$ to $F_3$ is a weighted average of the ratios of conformal block coefficients. From this we can immediately infer:
\benn
\textrm{there must exist at least one operator } \Ocal_i \textrm{ with } \fr{G^{\Ocal_i}_4}{G^{\Ocal_i}_3} \geq \fr{F_4^{\min}}{F_3^{\max}} \textrm{ and } \lambda_{\vep T \Ocal_i}^2 \neq 0.
\eenn
If there is more than one such operator, we clearly cannot place a lower bound on any individual OPE coefficient without imposing more constraints, at best we can bound a linear combination of their squared OPE coefficients. However, if there is a \emph{unique} operator whose ratio of conformal block coefficients satisfies this constraint, we can place a nonzero lower bound on its OPE coefficient from~\eqref{eq:F4F3Bound}.

In the 3d Ising CFT, this is the case if we make the (incorrect) simplifying assumption that only operators with $J=4$ contribute to $F_3$ and $F_4$. As we can see from Figure~\ref{fig:G3G4andG5G6}, the minimal-twist operator $C$ is the only spin-$4$ operator with a large enough ratio between conformal block coefficients,
\be
\fr{G^C_4}{G^C_3} = 49.4 \; > \; \fr{F_4^{\min}}{F_3^{\max}} = 5.64 \; > \; \fr{G^{C'}_4}{G^{C'}_3} = 4.02.
\ee
If spin-$4$ operators were the only contributions in~\eqref{eq:F4F3Bound}, we would therefore be able to infer that $\lambda_{\vep T C}^2 \neq 0$.

\begin{figure}[t!]
\centering
\includegraphics[width=.9\linewidth]{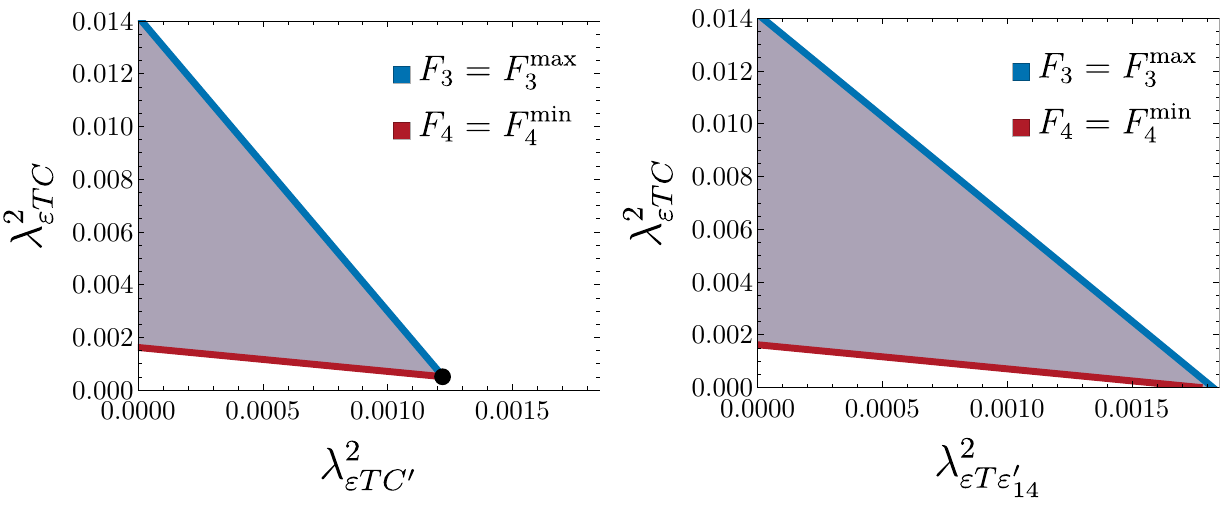}
\caption{\emph{Left:} Allowed OPE coefficients for $C$ and $C'$ satisfying the bounds on $F_3$ and $F_4$, when all other OPE coefficients are set to zero. The minimal allowed value for $\lambda_{\vep TC}^2$ is determined by the intersection of the two constraints (black dot). \emph{Right:} Allowed OPE coefficients for $C$ and the spin-$14$ operator $\vep'_{14}$ satisfying the bounds on $F_3$ and $F_4$, with all other OPE coefficients set to zero. The only lower bound is the trivial one $\lambda_{\vep TC}^2 \geq 0$.}
\label{fig:Min2D}
\end{figure}

In this simplified example with only spin-$4$ operators, how small could we make $\lambda_{\vep T C}^2$ while still satisfying~\eqref{eq:F4F3Bound}? It is straightforward to show that the minimal value is obtained by choosing only two OPE coefficients to be nonzero: the OPE coefficient for $C$ (the operator whose OPE coefficient we're trying to minimize) and the OPE coefficient for $C'$ (the operator with the second-largest value for $G^\Ocal_4/G^\Ocal_3$ after $C$). In the language of optimization, any other configuration would be wasting our finite $F_3$ ``budget'' on less ``efficient'' operators, therefore requiring a larger contribution from $C$ to ensure we attain our required $F_4$ ``goal''. We can therefore reduce our original problem to a two-dimensional one:
\be
\ba
\textrm{minimize} &&& \lambda_{\vep T C}^2 \\
\textrm{subject to} &&& \lambda_{\vep T C}^2 \; G^{C}_{3} + \lambda_{\vep T C'}^2 \; G^{C'}_{3} \leq F_{3}^{\max}, \\
&&& \lambda_{\vep T C}^2 \; G^{C}_{4} + \lambda_{\vep T C'}^2 \; G^{C'}_{4} \geq F_{4}^{\min}.
\ea
\label{eq:minCCp}
\ee
As we can see in the left plot in Figure~\ref{fig:Min2D}, these constraints restrict the allowed values of the two OPE coefficients to a finite area. The minimal value for $\lambda_{\vep T C}^2$ corresponds to the vertex at the intersection of the two constraints, giving us the lower bound (in this simplified example):
\be
\lambda_{\vep T C}^2 \geq 0.000506 \quad \ra \quad |\lambda_{\vep\hat{T}C}| \equiv \fr{|\lambda_{\vep T C}|}{\sqrt{c_T}} \geq 0.237.
\ee

Of course, spin-$4$ operators are not the only ones to contribute to $F_3$ and $F_4$, so to obtain an absolute lower bound on $\lambda_{\vep TC}^2$ we need to remove this simplifying assumption and consider operators of arbitrary spin. From Figure~\ref{fig:G3G4andG5G6}, we see that there are known operators with $J\geq 6$ whose conformal block coefficient ratios $G_4^\Ocal / G_3^\Ocal$ are larger than that of $C$. Including these operators in~\eqref{eq:F4F3Bound} therefore weakens the resulting lower bound on $\lambda_{\vep TC}^2$.

As we consider higher values of $J$, we eventually find known operators whose ratios $G_4^\Ocal / G_3^\Ocal$ are above the threshold set by $F_4^{\min} / F_3^{\max}$. The first such operator is $\vep'_{14}$, i.e.~the spin-$14$ operator with the second-lowest twist. If we replace $C'$ with $\vep'_{14}$ in the linear optimization problem~\ref{eq:minCCp}, we find that the OPE coefficients $\lambda_{\vep TC}^2$ and $\lambda_{\vep T\vep'_{14}}^2$ are limited to the finite region shown in the right plot of Figure~\ref{fig:Min2D}. However, unlike the case with $C$ and $C'$, the intersection of the two constraints is located \emph{below} $\lambda_{\vep TC}^2 = 0$, which means the only lower bound we can derive is the trivial one
\be
|\lambda_{\vep\hat{T}C}| \geq 0.
\ee

\begin{figure}[t!]
\centering
\includegraphics[width=.75\linewidth]{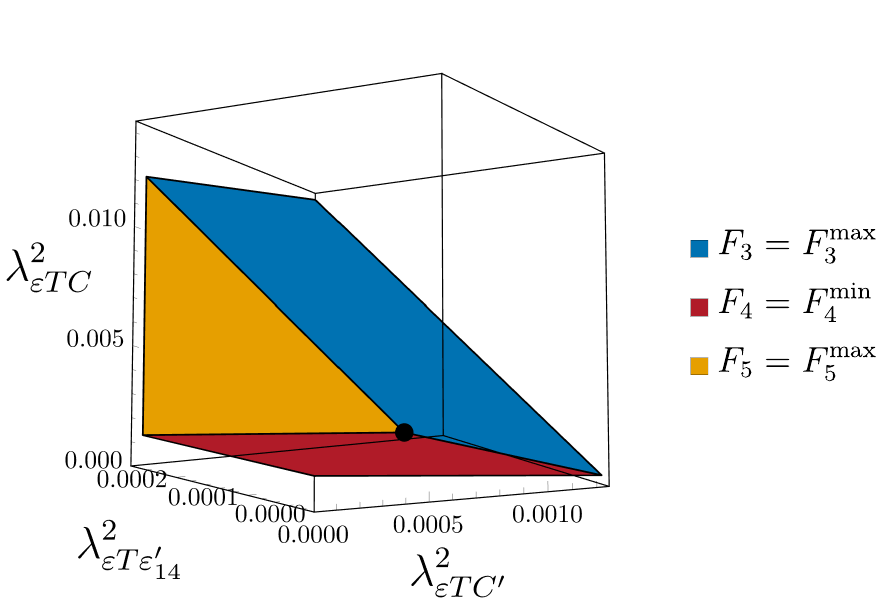}
\caption{Allowed OPE coefficients for $C$, $C'$, and the spin-$14$ operator $\vep'_{14}$ satisfying the bounds on $F_3$, $F_4$, and $F_5$, when all other OPE coefficients are set to zero. The minimal allowed value for $\lambda_{\vep TC}^2$ is determined by the intersection of the three constraints (black dot).}
\label{fig:Min3D}
\end{figure}

The situation is not hopeless, however, since so far we've only used the bounds on $F_3$ and $F_4$, ignoring all other $F_\ell$. From section~\ref{sec:IsingPartialWave}, we know that $F_5$ also has an upper bound, which in turn places an upper bound on the OPE coefficients of all operators with $J > 4$. To see how this new constraint can lead to a lower bound on $\lambda_{\vep TC}^2$, let's consider the three-dimensional linear optimization problem:
\be
\ba
\textrm{minimize} &&& \lambda_{\vep T C}^2 \\
\textrm{subject to} &&& \lambda_{\vep T C}^2 \; G^{C}_{3} + \lambda_{\vep T C'}^2 \; G^{C'}_{3} + \lambda_{\vep T \vep'_{14}}^2 \; G^{\vep'_{14}}_{3} \leq F_{3}^{\max}, \\
&&& \lambda_{\vep T C}^2 \; G^{C}_{4} + \lambda_{\vep T C'}^2 \; G^{C'}_{4} + \lambda_{\vep T \vep'_{14}}^2 \; G^{\vep'_{14}}_{4} \geq F_{4}^{\min}, \\
&&& \lambda_{\vep T \vep'_{14}}^2 \; G^{\vep'_{14}}_{5} \leq F_{5}^{\max},
\ea
\label{eq:minCCpE14}
\ee
where the last constraint only includes $\vep'_{14}$ because $G_5^C = G_5^{C'} = 0$. These constraints limit the three OPE coefficients to the interior of a polyhedron, as shown in Figure~\ref{fig:Min3D}, and the minimal value for $\lambda_{\vep T C}^2$ corresponds to the vertex at the intersection of all three constraints, giving us the lower bound:
\be
\lambda_{\vep T C}^2 \geq 0.000438 \quad \ra \quad |\lambda_{\vep\hat{T}C}| \geq 0.220.
\ee

\begin{figure}[t!]
\centering
\includegraphics[width=.9\linewidth]{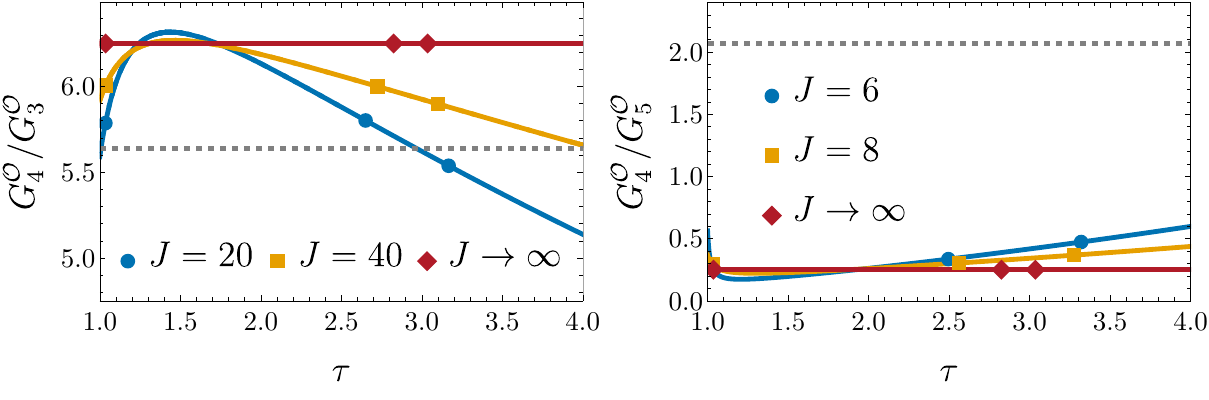}
\caption{\emph{Left:} Ratio of the $\ell=4$ conformal block coefficient $G^\Ocal_4$ to the $\ell=3$ conformal block coefficient $G^\Ocal_3$ as a function of the operator's twist $\tau \equiv \De-J$, for $J=20$ (blue), $J=40$ (yellow), and the limit $J\ra\infty$ (red). \emph{Right:} Ratio of $G^\Ocal_4$ to the $\ell=5$ conformal block coefficient $G^\Ocal_5$ as a function of $\tau$, for $J=6$ (blue), $J=8$ (yellow), and the limit $J\ra\infty$ (red). In both plots, the solid lines show the general dependence on $\tau$ and each marker corresponds to a known operator in the $\vep \times T$ OPE for the 3d Ising CFT. The dashed gray line indicates the threshold that at least one operator must satisfy for the energy correlator in the background of $\vep$ to be positive.}
\label{fig:G3G4G5}
\end{figure}

To find the absolute lower bound on $\lambda_{\vep TC}^2$, we need to repeat the linear optimization procedure~\eqref{eq:minCCpE14} for all possible combinations of $C$ with two other operators.\footnote{In general, the optimal configuration only involves the same number of variables as the number of constraints~\cite{Bertsimas:1997}.} While we of course don't know the scaling dimensions of all operators in the 3d Ising CFT, we can determine which operators lead to the optimal bound on $\lambda_{\vep TC}^2$ by comparing the ratios $G_4^\Ocal / G_3^\Ocal$ and $G_4^\Ocal / G_5^\Ocal$. Figure~\ref{fig:G3G4G5} shows these ratios for different values of $J$. As we can see, all low-twist operators with $J > 4$ have $G_4^\Ocal / G_5^\Ocal$ significantly below the threshold $F_4^{\min} / F_5^{\max}$, which means that these operators all have a strict upper bound on their OPE coefficient, leading to a non-trivial lower bound on $\lambda_{\vep TC}^2$.

The lowest possible bound comes from combining $C$ with $C'$ and the operator with the largest value for $G_4^\Ocal / G_3^\Ocal$, which as we can see from Figure~\ref{fig:G3G4G5} requires us to consider operators with $J \ra \infty$. The ratio of conformal block coefficients greatly simplifies in this limit, reducing to
\be
\lim_{J\ra\infty} \fr{G_\ell^\Ocal}{G_{\ell'}^\Ocal} = \fr{\G^2(\ell+d-1) \; \G^2(\ell'-1)}{\G^2(\ell'+d-1) \; \G^2(\ell-1)}.
\label{eq:RatioLimit}
\ee
Amazingly, this limiting value for the ratio is \emph{independent of the twist} of $\Ocal$, so we don't even need to know the large-$J$ spectrum to determine the resulting lower bound on $\lambda_{\vep TC}^2$. Repeating the linear optimization procedure~\eqref{eq:minCCpE14} with $C$, $C'$, and ``$\vep_\infty$'', we obtain the final lower bound~\eqref{eq:LowerBoundC}.


\bibliographystyle{utphys}
\bibliography{References}

\end{document}